\documentclass[fleqn,usenatbib]{mnras}
\usepackage{newtxtext,newtxmath}
\usepackage[T1]{fontenc}
\usepackage{ae,aecompl}
\usepackage{graphicx}	
\usepackage{amsmath}	
\usepackage{amssymb}	

\usepackage{pslatex}

\title[Project VeSElkA: HD~157087]{Project VeSElkA: Search for vertical stratification of element abundances in HD~157087\protect\thanks{Based on observations obtained at the Canada-France-Hawaii Telescope (CFHT) which is operated by the National Research Council of Canada, the Institut National des Sciences de l'Univers of the Centre National de la Recherche Scientifique of France, and the University of Hawaii. The operations at the Canada-France-Hawaii Telescope are conducted with care and respect from the summit of Maunakea which is a significant cultural and historic site.}}
\author[Khalack V.]
       {  V. Khalack$^1$ \\
          $^1$D\'epartement de Physique et d'Astronomie, Universit\'e de Moncton, Moncton, N.-B., Canada E1A 3E9\\
       }
\date{Accepted 27 February 2018;
      Received ???;
      in original form ???}

\pagerange{\pageref{firstpage}--\pageref{lastpage}} \pubyear{2017}

\begin{document}

\maketitle

\label{firstpage}

\begin{abstract}
The new spectropolarimetric spectra of HD~157087 obtained recently with ESPaDOnS at CFHT are analysed to verify the nature of this object. The fundamental stellar parameters $T_{\rm eff}$ = 8882 K, $\log{g}$=3.57 have been obtained for HD~157087 from the analysis of nine Balmer line profiles in two available spectra. Comparison of the results of our abundance analysis with the previously published data shows a variability of average abundance with time for some chemical species, while the abundance of other elements remains almost the same. The abundance analysis also reveals evidence of a significant abundance increase towards the deeper atmospheric layers for C, S, Ca, Sc, V, Cr, Mn, Co, Ni and Zr. Together with the found enhanced abundance of Ca and Sc this fact contradicts the classification of HD~157087 as a marginal Am star. Analysis of the available measurements of radial velocity results in detection of long periodic and short periodic variations. The long periodic variation supports the idea that HD~157087 is an astrometric binary system with a period higher than 6 years. The presence of short periodic variation of $V_{\rm r}$, as well as the detection of temporal variation of average abundance suggest that HD~157087 may be a triple system, where a short periodic binary rotates around a third star. In this case, the short periodic binary may consist of slowly rotating Am and A (or Ap with weak magnetic field) stars that have similar effective temperature and surface gravity, but different abundance peculiarities.
\end{abstract}

\begin{keywords}
atomic processes  -- line: profiles -- stars: atmospheres -- stars: chemically peculiar -- (stars:) binaries: general -- stars: individual: HD~157087
\end{keywords}

\section{Introduction}

Significant amount of the main sequence stars within the range of spectral classes B2-F4 show peculiarity of chemical abundance of different elements with respect to their solar abundance. These stars have been named as chemically peculiar (CP) stars and were initially classified by \citet{Preston74} in four distinct groups depending on the type of chemical peculiarity. More detailed classification of the CP stars has been proposed lately by \citet{Maitzen1984} and \citet{Smith1996}. The classification of \citet{Preston74} is widely used by authors to compile catalogues of know CP stars in both hemispheres \citep{Renson+91,Bychkov+03}. The most complete catalogue of the known and suspected Ap, HgMn and Am stars has been recently published by \citet{Renson+Manfroid09}.

The observed chemical peculiarity of CP stars can be explained in terms of atomic diffusion mechanism \citep{Michaud70}. In a hydrodynamically stable stellar atmosphere a competition between the gravitational and radiative forces can lead to accumulation or depletion of chemical elements at certain optical depths. An effective work of the atomic diffusion can result in a vertical stratification of abundance of chemical species \citep{HBH+00,LeBlanc+09}. A direct detection of vertical stratification of element abundances in stellar atmospheres of CP stars will argue in favor of an effective work of the atomic diffusion mechanism and will help to better understand the evolution of CP stars that results in the observed peculiarities of chemical abundance \citep{Khalack+LeBlanc15a}.

To search for the signs and to study the vertical stratification of element abundances in stellar atmospheres of CP stars of different types, our group has initiated project VeSElkA \citep{Khalack+LeBlanc15a,Khalack+LeBlanc15b}, which stands for Vertical Stratification of Element Abundances. A hydrodynamically stable atmosphere may be present in the main sequence stars of intermediate masses with a slow axial rotation. For this reason the only slowly rotating CP stars with V$\sin{i} <$ 40 km s$^{-1}$ were selected for the VeSElkA project from the catalogue 
of \citet{Renson+Manfroid09}. The CP stars with a small value of V$\sin{i}$ usually exhibit a number of sharp and mostly unblended line profiles well suitable for abundance analysis and for search of signs of vertical abundance stratification \citep{Khalack+13,Khalack+14,Khalack+17}.

Object HD~157087 (HR~6455, HIP~84821, BD+25$^{\circ}$3246) is mentioned by \citet{Renson+Manfroid09} as a candidate to CP stars of spectral class A2. Taking into account its relatively small rotational velocity $V \sin{i}$ = 15 km s$^{-1}$ \citep{Royer+02} it has been selected for the project VeSElkA. The first remarks about the chemical peculiarity of HD~157087 were published by \citet{Morgan32} who has detected the presence in its spectra of relatively strong line of Eu\,{\sc ii} 4205.05\AA. Until recently HD~157087 was considered as a poorly studied marginally Am star that shows some variability of radial velocity with time and probably is a spectroscopic binary \citep{Bidelman88}. Using the new DAO spectra of HD~157087 \citet{Yuce+11} have performed their detailed abundance analysis and detected a significant overabundance of chemical elements with Z $\geq$ 27. These authors also confirm the variability of radial velocity of this star with an amplitude around 8.6 km s$^{-1}$, but specify that more data is needed to find out correct values of orbital period and of radial velocity amplitude. \citet{Yuce+11} have confirmed classification of HD~157087 as an Am star following the description of Am stars given by \citet{Preston74} and \citet{Adelman87}, because its stellar atmosphere is rich in heavy elements and shows some deficit of Ca and Sc.

Am stars possess a convective stellar atmosphere \citep{Michaud+83} where turbulent motions rapidly decrease any abundance inhomogeneities (vertical or horizontal) leading to a more or less uniform abundance at different optical depths. The enrichment of certain chemical elements observed in the upper atmosphere of Am stars is delivered by convective motions from the bottom of convection zone, where those elements are overabundant due to the effect of atomic diffusion \citep{Michaud+83,Michaud+15}. If HD~157087 is indeed an Am star the abundance of chemical species can be different from the solar one \citep{Yuce+11}, but should be uniformly distributed in its stellar atmosphere.

In this study the new ESPaDOnS spectra of HD~157087 are thoroughly analysed with the aim to determine abundance of chemical elements 
and to verify the nature of this object that presumably is an Am star in a long period binary system. The observations and the procedure of spectra reduction are described in Section~\ref{obs}. The approaches used to determined the fundamental stellar parameters are discussed in Section~\ref{parameters}. The fitting procedure is described 
in Section~\ref{fit}, while the results of abundance analysis are presented in Section~\ref{analysis}. A binary nature of HD~157087 is considered in Section~\ref{binary} and
the discussion follows in Section~\ref{discus}.

\begin{table}
\begin{center}
\caption{Journal of spectropolarimetric observations and respective estimates of the mean longitudinal magnetic field in HD~157087 }
\label{tab1}
\begin{tabular}{ccrccc}\hline
 Date & HJD & $t_{exp}$ & S/N  & $<B_{\rm z}>$& $<N_{\rm z}>$\\
(UTC) &(2400000+)& (s)  &  Stokes I/V & (G) & (G) \\
\hline
 2014 Feb 10 & 56699.11730 &  764 & 820/650 & +40$\pm$36 & -20$\pm$36 \\
 2014 Feb 10 & 56699.12862 &  764 & 850/670 & -20$\pm$35 & +2$\pm$35 \\
 2014 Feb 15 & 56704.08525 &  764 & 900/700 & -11$\pm$33 & -4$\pm$33 \\
\hline
\end{tabular}
\end{center}
\end{table}

\section{Observations and data reduction}
\label{obs}

HD~157087 has been observed on February 10 and 15, 2014 with the spectropolarimetre ESPaDOnS (Echelle SpectroPolarimetric Device for Observations of Stars)
that works in the spectral domain from 3700\AA\, to 10000\AA\, employing the deep-depletion e2v device Olapa (see Table~\ref{tab1}).
The instrument performances and optical characteristics of the spectropolarimetre are described in detail by \citet{Donati+06}\footnote{More details about this instrument are available at {\rm www.cfht.hawaii.edu/Instruments/Spectroscopy/Espadons/}}. The high resolution (R=65000) Stokes I and V spectra with S/N$>$500 were acquired with ESPaDOnS and reduced employing the specially designed software Libre-ESpRIT \citep{Donati+97} which yields the circular polarisation spectrum (Stokes V) and the normal spectrum (Stokes I).

\begin{figure}
\begin{center}
\includegraphics[width=4.6in,angle=-90]{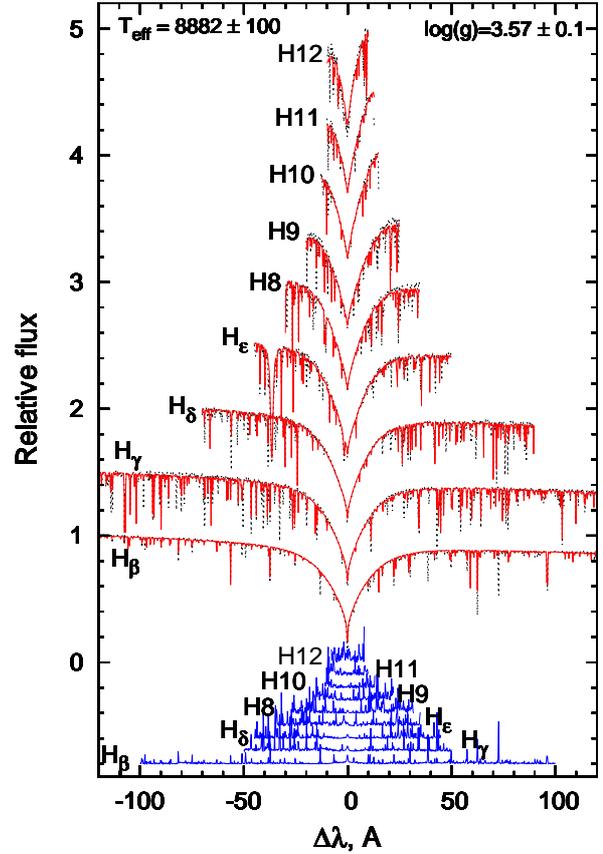} \\
\caption{The synthetic profiles (thin dotted line) obtained with the FITSB2 code for $T_{\rm eff}$ = 8882 K, $\log{g}$ = 3.57, [M/H]= 0.0 ($\chi^2/\nu$ = 0.5758) represent a relatively good approximation of the observed Balmer line profiles (thick line) for HD~157087.
In the bottom the differences between the observed and synthetic spectra are shown and shifted by 0.1 for visibility. The Balmer line profiles are shifted by 0.5 for the same reason. }
\label{fig1}
\end{center}
\end{figure}

\begin{table*}
\begin{center}
\caption{List of fundamental stellar parameters derived for HD~157087 from the analysis of Balmer line profiles employing grids of synthetic fluxes simulated with the PHOENIX15 and PHOENIX16 codes (see Subsection~\ref{Balmer_par}). }
\label{tab2}
\begin{tabular}{ccccrccc}\hline
PHOENIX& HJD & $T_{\rm eff}$, & $\log(g)$ & $V \sin{i}$ & $V_{\rm r}$& [M/H] &$\chi^2/\nu$\\
 &(2400000+)& K & & km s$^{-1}$ & km s$^{-1}$ & & \\
\hline
15 & 56699.11730 & 8873$\pm$100 & 3.56$\pm$0.10 & 10$\pm$2$^a$ & -11.0$\pm$ 0.5 & 0.00$\pm$0.10 & 0.6059 \\
16 & 56699.11730 & 8618$\pm$100 & 3.56$\pm$0.10 & 10$\pm$2$^a$ & -11.2$\pm$ 0.5 & 0.00$\pm$0.10 & 0.5460 \\
15 & 56699.12862 & 8881$\pm$100 & 3.57$\pm$0.10 & 10$\pm$2$^a$ &  -9.6$\pm$ 0.5 & 0.00$\pm$0.10 & 0.5928 \\
16 & 56699.12862 & 8672$\pm$100 & 3.57$\pm$0.10 & 10$\pm$2$^a$ & -11.4$\pm$ 0.5 & 0.00$\pm$0.10 & 0.5306 \\
15 & 56704.08525 & 8882$\pm$100 & 3.57$\pm$0.10 & 10$\pm$2$^a$ & -10.6$\pm$ 0.5 & 0.00$\pm$0.10 & 0.5758 \\
16 & 56704.08525 & 8629$\pm$100 & 3.56$\pm$0.10 & 10$\pm$2$^a$ & -10.6$\pm$ 0.5 & 0.00$\pm$0.10 & 0.5346 \\
\hline
\end{tabular}
\end{center}
{\it Notes:} $^a$From preliminary analysis of several Si\,{\sc ii} lines and from \citet{Abt+Morell95}.
\end{table*}

Table~\ref{tab1} presents a journal of spectropolarimetric observations of HD~157087, where the first and the second columns show the UTC date and the HJD of spectral acquisition. The exposure duration and the S/N ratio (in Stokes I and V spectra) are given respectively in the third and the fourth  columns. Estimates of the mean longitudinal magnetic field and the null field for this star are given in the fifth and the sixth columns respectively (see Subsec.~\ref{field}).

The effective temperature, surface gravity and metallicity have been determined from the best fit to the Balmer line profiles observed in the non-normalized spectra (see Section~\ref{parameters} and Fig.~\ref{fig1}). The reduced spectra were normalised to the continuum to study line profiles that belong to the helium and metals (see Section~\ref{analysis}).

\section{Estimation of fundamental stellar parameters}
\label{parameters}
The effective temperature and the surface gravity of HD~157087 were determined using two different methods for photometric temperature calibration and by fitting the non-normalised profiles of Balmer lines (except $H_{\rm \alpha}$) visible in the three available spectra of this star (see Fig.~\ref{fig1}).

\subsection{Analysis of Balmer line profiles}
\label{Balmer_par}

The Balmer lines profiles of HD~157087 have been fitted with the help of FITSB2 code \citep{Napiwotzki+04} using the grids of synthetic fluxes calculated for different $T_{\rm eff}$, $\log{g}$ and metallicity employing the code PHOENIX \citep{Hauschildt+97}. To derive a best fit parameters ($T_{\rm eff}$, $\log{g}$ and metallicity) of stellar atmosphere model the code FITSB2 performs fitting only of the Balmer line profiles. Nevertheless, during this procedure it takes into account some strong metallic lines that are present at the Balmer wings. FITSB2 code does not perform an abundance analysis for each chemical element, but it rather results in the estimate of metallicity which serves as an average measure of abundance of metals. For this reason, the spectral lines of most metals are not well fitted in Fig.~\ref{fig1} (see curves for the differences between the synthetic and observed spectra presented at the bottom of this image).

In this study we have used grids of synthetic fluxes calculated by \citet{Khalack+LeBlanc15a} with spectral resolution R=60000 for models with different metalicities using the version 15 of the code PHOENIX and grids of synthetic fluxes\footnote{The grids of synthetic spectra are available at http://phoenix.astro.physik.uni-goettingen.de/} simulated by \citet{Husser+13} for models with different metalicities and abundances of the $\alpha$-elements (O, Ne, Mg, Si, S, Ar, Ca, and Ti) employing the version 16 of the code PHOENIX \citep{H+B+99}. \citet{Husser+13} have calculated the grids of synthetic fluxes with spectral resolution R=500000, which we reduced to R=50000 when compiled those grids to the format read by the FITSB2 code.
Comparing to the PHOENIX15 code, its new version, 16, employs an updated list of atomic and molecular lines and uses a new equation of state to simulate a model of stellar atmosphere and respective synthetic flux \citep{Husser+13}.

The fitting results obtained for the three available spectra (see Table~\ref{tab1}) of HD~157087 are shown in the Table~\ref{tab2}. In all cases the fitting procedure is carried out assuming a rotational velocity $V \sin{i}$ = 10 km s$^{-1}$ derived for HD~157087 from the preliminary analysis of Si\,{\sc ii} line profiles. For the three available spectra the fitting procedure that employs the same type of grids of synthetic fluxes (for example, calculated with the PHOENIX15 code) results in almost the same values of $T_{\rm eff}$, $\log{g}$ and metallicity (see Table~\ref{tab2}). The effective temperatures provided by the use of the PHOENIX16 grids of synthetic fluxes \citep{Husser+13} appear to be always significantly smaller that the $T_{\rm eff}$ obtained employing the PHOENIX15 grids of synthetic fluxes \citep{Khalack+LeBlanc15a}. Meanwhile, both types of grids result in the same values of the surface gravity and metallicity and in the similar values of the radial velocity. Fig.~\ref{fig1} presents an example of the best fit of Balmer lines profiles obtained using the grids of synthetic fluxes simulated with the PHOENIX15 code \citep{Khalack+LeBlanc15a}. The differences between the synthetic and observed spectra presented at the bottom of this image show that the obtained best fit approximates the analysed Balmer lines profiles relatively well even taking into account the contamination of Balmer wings by some metals lines.

\begin{figure*}
\begin{center}
\begin{tabular}{cc}
\includegraphics[width=2.1in,angle=-90]{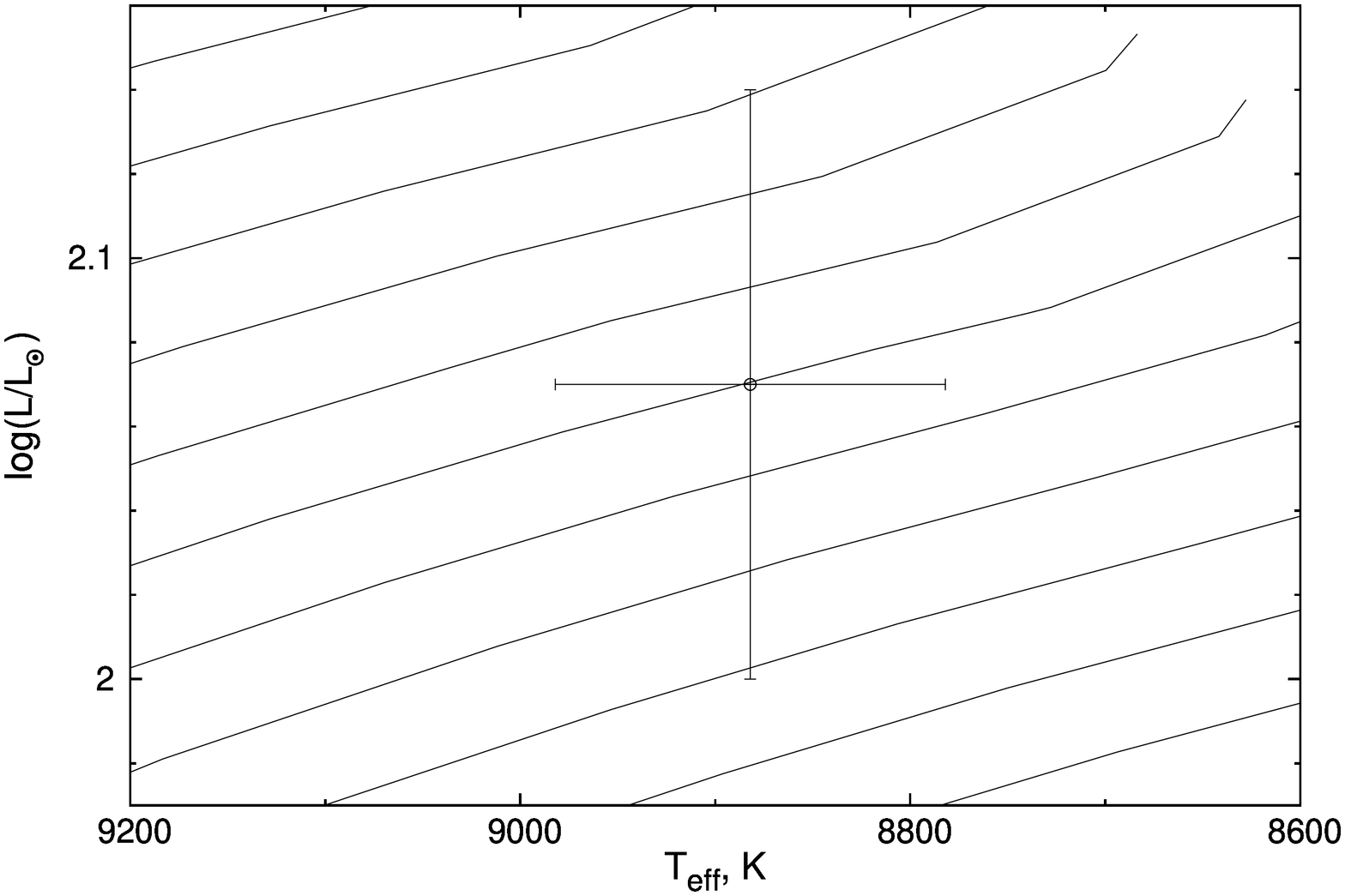} &
\includegraphics[width=2.1in,angle=-90]{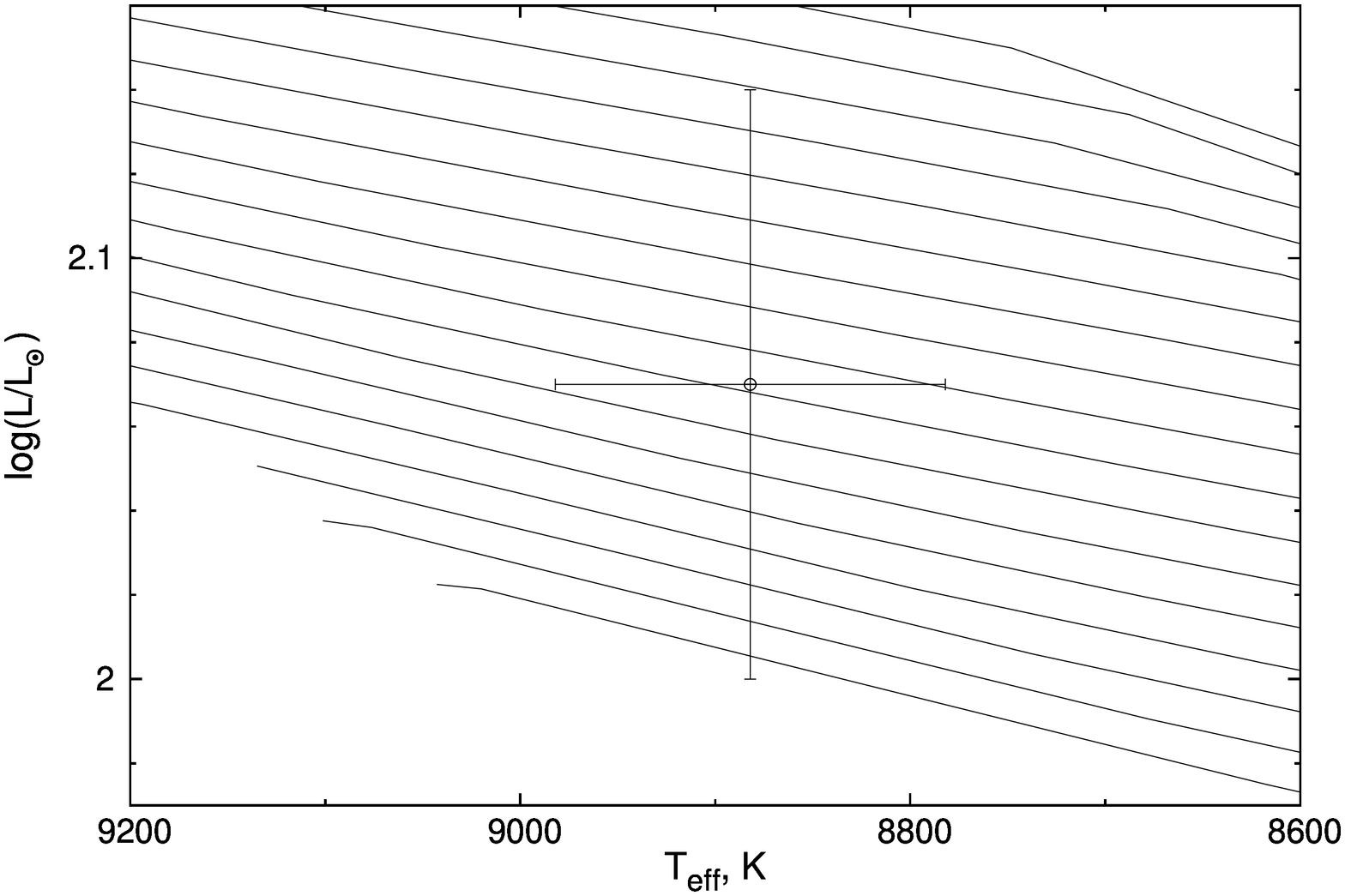} \\
\end{tabular}
\caption{ Position of HD~157087 in the Hertzsprung-Russell diagram that corresponds to the age $\log(t)$=8.64 (left) and $\log(t)$=8.73 (right). The constraints for $T_{\rm eff}$ and $\log(L/L_{\odot})$ are shown by the 1$\sigma$-error bars. Continuous lines represent isochrones calculated for the solar metallicity} [M/H]= 0.0.
\label{fig2}
\end{center}
\end{figure*}

\subsection{Photometric temperature calibrations}
\label{Photometric_T}

The effective temperature $T_{\rm eff}$= 8897$\pm$51~K has been derived for HD~157087 from the photometric temperature $[c_{\rm 1}]$-calibration introduced by \citet{Napiwotzky+93} for Ap stars. For this procedure we have used the photometric index $[c_{\rm 1}]$=1.177$\pm$0.012 determined according to \citet{Stromgren66} from the $uvby\beta$ photoelectric photometry of this star \citep{Napiwotzky+93}.

Taking into account the parallax measured by Hipparcos \citep{ESA97,vanLeeuwen07} the distance to HD~157087 is approximately 137~pc. The service for the 3D dust mapping\footnote{3D mapping of dust distribution in the solar neighborhood is available at http://argonaut.rc.fas.harvard.edu/query} provides for this object an interstellar reddening $\rm E(B-V)_{SFD}\approx$0.006 mag \citep{Schlafly+14,Green+14,Green+15} that according to \citet{Schlafly+Finkbeiner11} corresponds to $\rm E(B-V)\approx$0.0053 mag. Respectively, the colour $\rm B-V\approx$ 0.063 mag provided for HD~157087 by Hipparcos \citep{ESA97} is reduced to $\rm (B-V)_{0}\approx$0.058 mag and the employed photometric temperature calibration \citep{Netopil+08} results for this star in the effective temperature $T_{\rm eff}$= 8930$\pm$165~K.

Final results for $T_{\rm eff}$ are presented in Table~\ref{tab3}, where all three different methods result in similar values for the effective temperature taking into account their estimation errors.

\begin{table}
\begin{center}
\caption{Stellar parameters of HD~157087 derived from the spectral and photometric data. }
\label{tab3}
\begin{tabular}{lcc}\hline
Parameter & \multicolumn{2}{c}{Value} \\
\hline
Distance$^a$ (pc) & 136.99$\pm$0.08 & \\
Age: $\log(t)$ & 8.64$\pm$0.03 & 8.73$\pm$0.07  \\
$M_{\rm V}$ (mag) & -0.34$\pm$0.18 & \\
$\log(L_{\rm *}/L_{\rm \odot)}$ & 2.07$\pm$0.07 & \\
$T_{\rm eff}^b$ (K) & 8882$\pm$100 & \\
$R^f_{\rm *}/R_{\rm \odot}$ & 4.63$\pm$0.09  & 4.57$\pm$0.12  \\
$R^e_{\rm *}/R_{\rm \odot}$ & 4.53$\pm$0.09  & 4.42$\pm$0.12  \\
$M_{\rm *}/M_{\rm \odot}$ & 2.78$\pm$0.10 & 2.64$\pm$0.14 \\
$\log(g)$ & 3.55$\pm$0.10 & 3.54$\pm$0.10 \\
$T_{\rm eff}^c$ (K) & 8897$\pm$51 & \\
$T_{\rm eff}^d$ (K) & 8930$\pm$165 &  \\
\hline
\end{tabular}
\end{center}
{\it Notes:} $^a$data from Hipparcos \citep{vanLeeuwen07}, $^b$results derived from the fitting of Balmer lines using the grid of synthetic fluxes calculated with PHOENIX15 \citep{Hauschildt+97}, $^c$results derived from the $[c_{\rm 1}]$-photometric temperature calibrations \citep{Napiwotzky+93}, $^d$results obtained from the $\rm (B-V)_{0}$ photometric temperature calibrations \citep{Netopil+08}, $^f$radii derived from mass and $\log(g)$ known from the interpolation of isochrones \citep{Bressan+12}, $^e$radii obtained from known mass and $\log(g)$ derived from the fitting of Balmer lines.
\end{table}

\subsection{Age, mass, $\log{g}$ and radius}
\label{age}

Using the the known value of $\rm E(B-V)_{SFD}$ we can determine an interstellar extinction for HD~157087 \citep{Schlafly+Finkbeiner11} and estimate its luminosity and absolute magnitude based on the visual magnitude and parallax provided for this object by Hipparcos catalogue \citep{ESA97,vanLeeuwen07}. The distance, absolute magnitude and luminosity derived in this way are presented in Table~\ref{tab3}. Taking into account the effective temperature obtained from the analysis of Balmer line profiles, the aforementioned luminosity and their estimation errors one can locate the position of HD~157087 on the Hertzsprung-Russell (HR) diagram and determine its age, mass and surface gravity throughout interpolation between different isochrones\footnote{Isochrones for stars with different metallicities can be downloaded at {\rm stev.oapd.inaf.it/cgi-bin/cmd}.} calculated by \citet{Bressan+12} for the metallicity [M/H]=0.0 (see Fig.~\ref{fig2}). In this case the same values of $T_{\rm eff}$ and luminosity correspond to two different ages for which the proper stellar mass and surface gravity were estimated from the interpolation of isochrones (see Table~\ref{tab3}).
The surface gravity derived for the two different ages is in a good accordance with the $\log(g)$ obtained for HD~157087 from the fitting of Balmer line profiles taking into account the estimation errors (see Tables~\ref{tab2} and~\ref{tab3}).

Unfortunately, the data calculated by \citet{Bressan+12} for different isohrones do not include a direct information about the size of stellar radius. Nevertheless, one can estimate its maximal value from known stellar mass and surface gravity assuming an uniform distribution of mass in a sphere of stellar radius (see Table~\ref{tab3}). If we use the masses obtained from the interpolation of isochrones and the $\log(g)$ derived from the simulation of Balmer line profiles the estimates of stellar radius will be smaller by few percents.

\subsection{Magnetic field measurement}
\label{field}

Estimate of the mean longitudinal magnetic field (averaged over the visible stellar disk) has been deduced from the analysis of Stokes I and V profiles of $H_{\rm \alpha}$ line core using the approach that was initially developed by \citet{Landstreet+15}.
It has been tested employing the spectra-polarimetric observations (Stokes I and V spectra) of HD~16605 and HD~49299 and the derived measurements of the mean longitudinal magnetic field are  are similar to the results obtained by \citet{Landstreet+15} for these stars employing the Least Square Deconvolution method \citep{Landstreet+08}. The estimates of the mean longitudinal magnetic field $<B_{\rm z}>$ and of the null field $<N_{\rm z}>$ derived for HD~157087 are shown in Table~\ref{tab1}.

\subsection{Model of stellar atmosphere}
\label{model}

To carry out an abundance analysis we have simulated a model of stellar atmosphere for HD~157087 with the help of PHOENIX15 code \citep{Hauschildt+97} using the values $T_{\rm eff}$ = 8882~K and $\log(g)$ = 3.57 derived from the fitting of Balmer line profiles. Stellar radius $R_{\rm *}$ = 4.5 $R_{\rm \odot}$ provided for this stars by the Catalogue of Apparent Diameters and Absolute Radii \citep{Pasinetti-Fracassini+01} and measured by \citet{Wesselink+72} is also used to calculate model of stellar atmosphere. This radius is just slightly smaller than the radii obtained using the $\log(g)$ derived from the interpolation of isochrones, but it is similar to those radii obtained using the $\log(g)$ derived from the simulation of Balmer line profiles (see Table~\ref{tab3}). All the derived values of stellar radius indicate that HD~157087 has a luminosity class III \citep{Wesselink+72,Pasinetti-Fracassini+01}. Respectively, in the procedure to calculate the stellar atmosphere model we have employed approach where the surface gravity depends on the optical depth of modeled atmosphere \citep{Hauschildt+97} and assumed a microturbulence velocity $\xi$ = 2.1 km s$^{-1}$ \citep{Yuce+11}.

\section{Fitting procedure}
\label{fit}

During a preliminary check of normalised spectra we have selected for analysis the spectral regions with well visible continuum and a good S/N ratio, which are clean from telluric lines and are outside the wings of Balmer lines. Afterwards, an automatic procedure scans these regions for spectral line profiles (blended or unblended with a width less than 4\AA) that are deep enough for abundance analysis taking into account the local S/N ratio. Observational data for each selected line profile are stored in a separate file. This file is linked to a list of spectral lines (and respective atomic data) that belong to different ions of specified chemical species and contribute to formation of the selected profile.
This automatic procedure also creates a file with the input data for ZEEMAN2 code \citep{Landstreet88} and an HTML-file with the images of all line profiles selected for analysis. A user can easily verify the quality of all selected line profiles by looking at their images in any HTML-browser, make corrections and rerun the automatic procedure if necessary.

\begin{figure}
\includegraphics[width=1.65in,angle=-90]{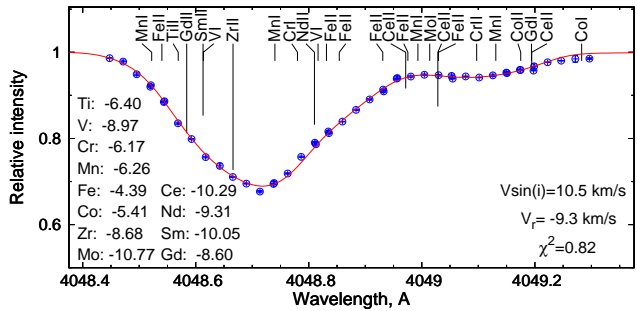}
\caption{The best fit of HD~157087 spectrum observed on February 10 (open circles) by a synthetic profile (solid line) calculated assuming the given (on the left side of the graph) abundance of chemical elements and the values of radial velocity and V$\sin{i}$ (presented at the bottom right corner) results in a relatively good fitting quality ($\chi^2/\nu$). The simulation has been carried out employing the stellar atmosphere model with $T_{\rm eff}$ = 8882K, $\log{g}$ = 3.57 (see Table~\ref{tab2}).
Vertical bars show position of spectral lines that contribute to this profile and their length is inversely proportional to the optical depths $\tau_{\rm 5000}$ of line-core formation (longer bars correspond to smaller optical depth). }
\label{fig3}
\end{figure}

The modified version of ZEEMAN2 code \citep{Khalack+Wade06,Khalack+07} has been used to automatically treat a list of a thousand line profiles in a consecutive mode. From the analysis of each line profile the user can determine an abundance of chemical elements that contribute to it, V$\sin{i}$ and radial velocity of a studied star (see Fig.~\ref{fig3}).
If a line profile is due to the contribution from only one spectral line of a certain ion, we can assume that the core of the line profile is formed mainly at line optical depth $\tau_{\rm \ell}$=1, which corresponds to a particular layer of the stellar atmosphere specified by a particular value of optical depth $\tau_{\rm 5000}$.
Taking into account that the different spectral lines usually have different oscillator strengths and lower energetic levels, their cores are generally formed at different optical depths $\tau_{\rm 5000}$. Analysis of a number of line profiles that belong to the same ion results in a certain distribution of ion abundance with optical depths.

In the case of a well mixed stellar atmosphere, one will derive a uniform distribution of ion abundance for all studied optical depths. Nevertheless, in a hydrodynamically stable atmosphere the atomic diffusion may lead to a vertical stratification of ion abundance \citep{Michaud70} and from the analysis of line profiles one can detect a certain pattern in the abundance change with optical depths. This method has been previously employed by \citet{Khalack+07,Khalack+08,Khalack+10,Khalack+13,Khalack+14,Khalack+17}, \citet{Thiam+10}, \citet{LeBlanc+15} and \citet{Ndiaye+17} to search for proofs of vertical abundance stratification of different chemical elements in stellar atmospheres of some BHB, post-HB and CP stars. Similar approach has been used recently by \citet{Castelli+17} to study the abundance peculiarities in HR~6000 (non-magnetic Bp stars), but employing instead the average optical depth of line profile formation \citep{Castelli05} to determine the respective value of $\tau_{\rm 5000}$.

\begin{table}
\centering
\label{tab4}
\caption{List of spectral lines and respective atomic data used to study the abundance peculiarities in HD~157087.
The complete lists of the spectral line profiles employed to carry out the abundance analysis of both spectra are available online.}
\begin{tabular}{llllc}
\hline
Ion & $\lambda$, \AA& $\log{N_{\rm ion}/N_{\rm tot}}$& $\log gf$ & $E_{\rm l}$, $cm^{-1}$ \\
\hline
C\,{\sc i} &  4932.049 & -4.314$\pm$0.012 & -1.658 &  61981.82  \\
C\,{\sc i} &  5380.337 & -4.452$\pm$0.002 & -1.616 &  61981.82  \\
C\,{\sc i} &  9061.440 & -3.734$\pm$0.004 & -0.347 &  60352.63  \\
C\,{\sc i} &  9088.510 & -4.141$\pm$0.005 & -0.430 &  60352.64  \\
\hline
\end{tabular}
\end{table}

Nevertheless, the observed line profiles even in sharp lined (because of a small value of V$\sin{i}$) CP stars can be blended to different degree due to contribution from ions of different chemical species (see, for example, Fig.~\ref{fig3}). In the approach used in this study, for each selected line profile we have compiled a list of spectral lines that contribute to the studied profile. If the contribution of a particular ion to the formation of this profile is done trough only one blend (line), we determine the optical depth of core formation for this line ($\tau_{\rm \ell}$=1) and after derive the respective value of $\tau_{\rm 5000}$. In the case, that several blends (lines) that belong to the same ion are present in the line list, we determine the optical depth of core formation of the strongest component and then derive the respective optical depth $\tau_{\rm 5000}$. This procedure is repeated for all the different ions that contribute to the studied profile and whose abundance is determined from the fitting procedure. Analysis of a thousand line profiles will results in a pattern of abundance distribution with optical depth for all studied ions if they are represented by a number of spectral lines in the observed stellar spectra (see, for example, Fig.~\ref{fig5}).

To carry out the abundance analysis of the available spectra of HD~157087 we have employed lines and respective atomic data \citep{Kramida+15,Kupka+00,Ryab+15} of the first two ionisation states of 39 chemical species 
to properly fit the selected line profiles with the help of ZEEMAN2 code \citep{Landstreet88,Khalack+Wade06,Khalack+07}.
An example of the list of spectral lines used for the abundance analysis is shown in Table~\ref{tab4} whose full versions is available online for both studied spectra.
Taking into account the high resolution of HD~157087 spectra obtained with ESPaDOnS (see Section~\ref{obs}), the number of spectral measurements (data at certain wavelengths) in a studied line profile always exceeds the amount of free model parameters used to fit this profile with a synthetic one (see, for example, Fig.~\ref{fig3}). Therefore one can estimate the values of radial ($V_{\rm r}$) and rotational ($V \sin{i}$) velocities, and of element's abundance with a relatively good precision. Taking into account the given S/N ratio of the analysed spectra (see Table~\ref{tab1}) the precision of abundance estimates is not high for the elements whose lines contribute to the formation of weak line profiles or of weak blends in strong profiles. In both these cases, one deals with the lines formed in the deep layers of stellar atmosphere and should pay a special attention to the derived abundance estimates. Nevertheless, a combination of these estimates with the results derived for the same ion from the analysis of other line profiles helps one to recreate a general tendency of its abundance variation with optical depth. For this reason the spectra obtained with a high spectral resolution and a high signal-to-noise ratio are most suitable for the abundance analysis that employs the aforementioned approach to search for signatures of vertical stratification of element's abundance.

\section{Abundance analysis}
\label{analysis}

Taking into account that the derived estimates of the longitudinal magnetic field are negligibly small for all analysed spectra (see Subsection~\ref{field} and Table~\ref{tab1}), only Stokes I spectra have been used in this study to carry out an abundance analysis for HD~157087. 
The model of stellar atmosphere calculated for $T_{\rm eff}$ = 8882 K, $\log{g}$ = 3.57, [M/H]= 0.0 and microturbulence velocity $\xi$ = 2.1 km s$^{-1}$ \citep{Yuce+11} with the help of PHOENIX15 code \citep{Hauschildt+97} was employed to simulate the synthetic line profiles. Each line profile selected for analysis has been fitted 4--5 times with the help of ZEEMAN2 code \citep{Landstreet88,Khalack+Wade06,Khalack+07} using a different set of initial values for average abundance of chemical elements, V$\sin{i}$ and radial velocity, that are fed to the fitting routine. This approach allows one to verify if the absolute minimum of $\chi^2$-function is reached and the best fit result are obtained. A post-simulation automatic procedure compares the "best fit data" from different simulations of the same line profile and select the best results with the smallest derived value of $\chi^2$-function.

\begin{table}
\begin{center}
\caption{Comparison of the averages abundances derived for HD~157087.}
\def\arraystretch{0.96}
\setlength{\tabcolsep}{3pt}
\begin{tabular}{l|cr|cr|lr}
\hline
   & \multicolumn{2}{c}{ Feb. 10$^a$} 
   & \multicolumn{2}{c}{ Feb. 15$^a$} 
   & \multicolumn{2}{c}{ \citet{Yuce+11} } \\
   & \multicolumn{4}{c}{$T_{\rm eff}$=8882K, $\log{g}$=3.57}& \multicolumn{2}{c}{ $T_{\rm eff}$=8700K, $\log{g}$=3.25 } \\
Ion & [X/H] & N & [X/H] & N & [X/H] & N \\
\hline
He\,{\sc i}  & -0.08$\pm$0.13 &  4 & -0.15$\pm$0.09 &   7 &                &      \\
C\,{\sc i}   & -0.33$\pm$0.08 & 26 & -0.25$\pm$0.08 &  19 & -0.14$\pm$0.13 &   3  \\
C\,{\sc ii}  & +0.09$\pm$0.10 &  4 &                &     &                &      \\
N\,{\sc i}   & -0.04$\pm$0.11 &  9 & -0.06$\pm$0.09 &  10 &            &      \\
N\,{\sc ii}  & -0.15$\pm$0.08 &  6 & +0.09$\pm$0.10 &   8 &             &      \\
O\,{\sc i}   & +0.02$\pm$0.07 & 16 & -0.06$\pm$0.09 &  13 &  -0.87   &   1  \\
O\,{\sc ii}  & +0.02$\pm$0.07 & 14 & +0.01$\pm$0.12 &   9 &                &      \\
Ne\,{\sc ii} & +0.02$\pm$0.11 & 52 & +0.05$\pm$0.11 &  39 &                &      \\
Na\,{\sc i}  & +0.73$\pm$0.12 &  7 & +0.25$\pm$0.50 &   1 &                &      \\
Mg\,{\sc i}  & +0.12$\pm$0.09 &  6 & -0.04$\pm$0.04 &   4 & -0.47$\pm$0.24 &   2  \\
Mg\,{\sc ii} & -0.20$\pm$0.06 &  9 & -0.22$\pm$0.07 &   9 & -0.07$\pm$0.23 &   4  \\
Al\,{\sc ii} & +1.38$\pm$0.41 &  4 & +1.11$\pm$0.21 &   7 & +0.27    &   1  \\
Si\,{\sc i}  & +0.12$\pm$0.10 & 20 & +0.19$\pm$0.12 &  19 &                &      \\
Si\,{\sc ii} & +0.17$\pm$0.11 & 18 & +0.24$\pm$0.09 &  19 & -0.01$\pm$0.19 &   9  \\
P\,{\sc i}   & +1.10$\pm$0.37 &  2 & +1.19$\pm$0.44 &   3 &                &      \\
P\,{\sc ii}  & +0.33$\pm$0.19 & 10 & +1.39$\pm$0.12 &   4 &                &      \\
S\,{\sc i}   & +0.37$\pm$0.06 & 12 & +0.30$\pm$0.10 &   9 & +0.38$\pm$0.12 &   3  \\
S\,{\sc ii}  & +0.79$\pm$0.10 &  9 & +0.70$\pm$0.12 &  12 &                &      \\
Ar\,{\sc ii} & +0.38$\pm$0.14 & 42 & +0.38$\pm$0.15 &  25 &                &      \\
K\,{\sc ii}  & +0.47$\pm$0.52 &  2 & +0.84$\pm$0.59 &   2 &                &      \\
Ca\,{\sc i}  & +0.22$\pm$0.08 & 38 & +0.12$\pm$0.07 &  29 & -0.12$\pm$0.16 &   8  \\
Ca\,{\sc ii} & +0.12$\pm$0.10 & 15 & -0.01$\pm$0.06 &  10 &                &      \\
Sc\,{\sc i}  & +0.90$\pm$0.23 &  7 & +1.78$\pm$0.19 &  13 &                &      \\
Sc\,{\sc ii} & +0.00$\pm$0.09 & 17 & +0.74$\pm$0.26 &   9 & -0.35$\pm$0.19 &   9  \\
Ti\,{\sc i}  & +0.99$\pm$0.10 & 38 & +0.90$\pm$0.10 &  36 & +0.30$\pm$0.08 &   6  \\
Ti\,{\sc ii} & +0.39$\pm$0.06 &109 & +0.34$\pm$0.06 &  96 & -0.14$\pm$0.20 &  51  \\
V\,{\sc i}   & +0.62$\pm$0.14 & 24 & +1.00$\pm$0.14 &  22 & +0.18    &   1  \\
V\,{\sc ii}  & +0.40$\pm$0.10 & 49 & +0.52$\pm$0.10 &  48 & +0.45$\pm$0.22 &  28  \\
Cr\,{\sc i}  & +0.86$\pm$0.07 &131 & +0.83$\pm$0.07 & 116 & +0.31$\pm$0.20 &  13  \\
Cr\,{\sc ii} & +0.53$\pm$0.07 &131 & +0.47$\pm$0.07 & 108 & +0.10$\pm$0.21 &  40  \\
Mn\,{\sc i}  & +0.75$\pm$0.10 & 62 & +0.70$\pm$0.08 &  74 & +0.11$\pm$0.16 &  20  \\
Mn\,{\sc ii} & +0.63$\pm$0.10 & 55 & +0.52$\pm$0.09 &  55 & +0.29$\pm$0.21 &  14  \\
Fe\,{\sc i}  & +0.20$\pm$0.04 &440 & +0.17$\pm$0.04 & 366 & +0.14$\pm$0.18 & 247  \\
Fe\,{\sc ii} & +0.41$\pm$0.05 &252 & +0.38$\pm$0.05 & 195 & +0.15$\pm$0.20 &  73  \\
Co\,{\sc i}  & +1.53$\pm$0.09 & 91 & +1.52$\pm$0.10 &  74 & +0.44$\pm$0.21 &   7  \\
Co\,{\sc ii} & +1.74$\pm$0.12 & 36 & +1.36$\pm$0.16 &  26 &                &      \\
Ni\,{\sc i}  & +0.59$\pm$0.05 &127 & +0.60$\pm$0.05 & 104 & +0.55$\pm$0.19 &  34  \\
Ni\,{\sc ii} & +0.91$\pm$0.10 & 33 & +0.80$\pm$0.08 &  28 & +0.65$\pm$0.16 &   6  \\
Cu\,{\sc ii} & -0.10$\pm$0.08 & 39 & -0.15$\pm$0.08 &  28 &                &      \\
Zn\,{\sc i}  & +0.64$\pm$0.07 &  3 & +0.74$\pm$0.06 &   4 & +0.67$\pm$0.08 &   3  \\
Sr\,{\sc ii} & +0.81$\pm$0.15 &  4 & +1.41$\pm$0.30 &   5 & +0.85$\pm$0.07 &   2  \\
Y\,{\sc i}   & +1.05$\pm$0.07 &  3 &                &     &                &      \\
Y\,{\sc ii}  & +0.82$\pm$0.06 & 33 & +0.82$\pm$0.07 &  28 & +0.79$\pm$0.20 &  14  \\
Zr\,{\sc i}  & +2.14$\pm$0.10 &  3 & +2.28$\pm$0.24 &   2 &                &      \\
Zr\,{\sc ii} & +1.00$\pm$0.05 & 43 & +1.02$\pm$0.05 &  44 & +0.95$\pm$0.18 &  32  \\
Mo\,{\sc i}  & -0.12$\pm$0.11 & 20 & -0.30$\pm$0.10 &  12 &                &      \\
Ba\,{\sc i}  & +3.00$\pm$0.09 &  2 &      &               &                &     \\
Ba\,{\sc ii} & +1.49$\pm$0.23 &  9 & +1.24$\pm$0.14 &   7 & +0.46$\pm$0.08 &   2  \\
La\,{\sc i}  &                &    & +0.99$\pm$0.50 &   1 &                &      \\
La\,{\sc ii} & +1.06$\pm$0.11 & 30 & +0.98$\pm$0.11 &  20 & +0.93$\pm$0.11 &  10  \\
Ce\,{\sc i}  & +1.41$\pm$0.19 &  4 & +1.44$\pm$0.50 &   1 &                &      \\
Ce\,{\sc ii} & +1.67$\pm$0.06 &199 & +1.58$\pm$0.07 & 162 & +0.97$\pm$0.18 &  31  \\
Pr\,{\sc ii} & +2.30$\pm$0.10 & 49 & +2.13$\pm$0.13 &  32 &                &      \\
Nd\,{\sc ii} & +1.69$\pm$0.08 & 77 & +1.52$\pm$0.08 &  84 & +0.84$\pm$0.07 &   3  \\
Sm\,{\sc i}  & +1.27$\pm$0.34 &  6 & +1.46$\pm$1.22 &   2 &                &      \\
Sm\,{\sc ii} & +1.87$\pm$0.13 & 36 & +1.53$\pm$0.12 &  43 &                &      \\
Eu\,{\sc ii} & +1.06$\pm$0.26 &  4 & +0.79$\pm$0.18 &   5 & +0.83$\pm$0.05 &   2  \\
Gd\,{\sc i}  & +0.86$\pm$0.33 &  5 & +1.31$\pm$1.01 &   2 &                &      \\
Gd\,{\sc ii} & +1.48$\pm$0.11 & 39 & +1.47$\pm$0.11 &  37 & +1.08$\pm$0.17 &   3  \\
Dy\,{\sc i}  & +1.25$\pm$0.06 &  3 & +1.10$\pm$0.50 &   1 &                &      \\
Dy\,{\sc ii} & +1.54$\pm$0.13 & 21 & +1.48$\pm$0.15 &  23 & +0.81$\pm$0.15 &   6  \\
Yb\,{\sc ii} & +0.17$\pm$0.18 &  8 & +0.71$\pm$0.24 &   7 &                &      \\
\hline
\end{tabular}
\label{tab5}
\end{center}
{\it Notes:} $^a$results for two spectra observed on different nights (see Table~\ref{tab1}).
\end{table}

Following the procedure developed by \cite{Khalack+17}, an upper cut-off limit has been imposed for the $\chi^2$-function that corresponds to a relatively "good fit". This limit depends on the S/N-ratio in the studied spectrum and is required to cut out the abundance analysis data gathered from the line profiles with a low fit quality.
The same procedure verifies each well fitted profile (with acceptable value of $\chi^2$-function) if the obtained data for the abundance estimates, for the radial velocity and V$\sin{i}$ are significantly different from their average values, and removes that data 
from the abundance analysis.

\begin{figure*}
\begin{center}
\begin{tabular}{cc}
\includegraphics[width=2.3in,angle=-90]{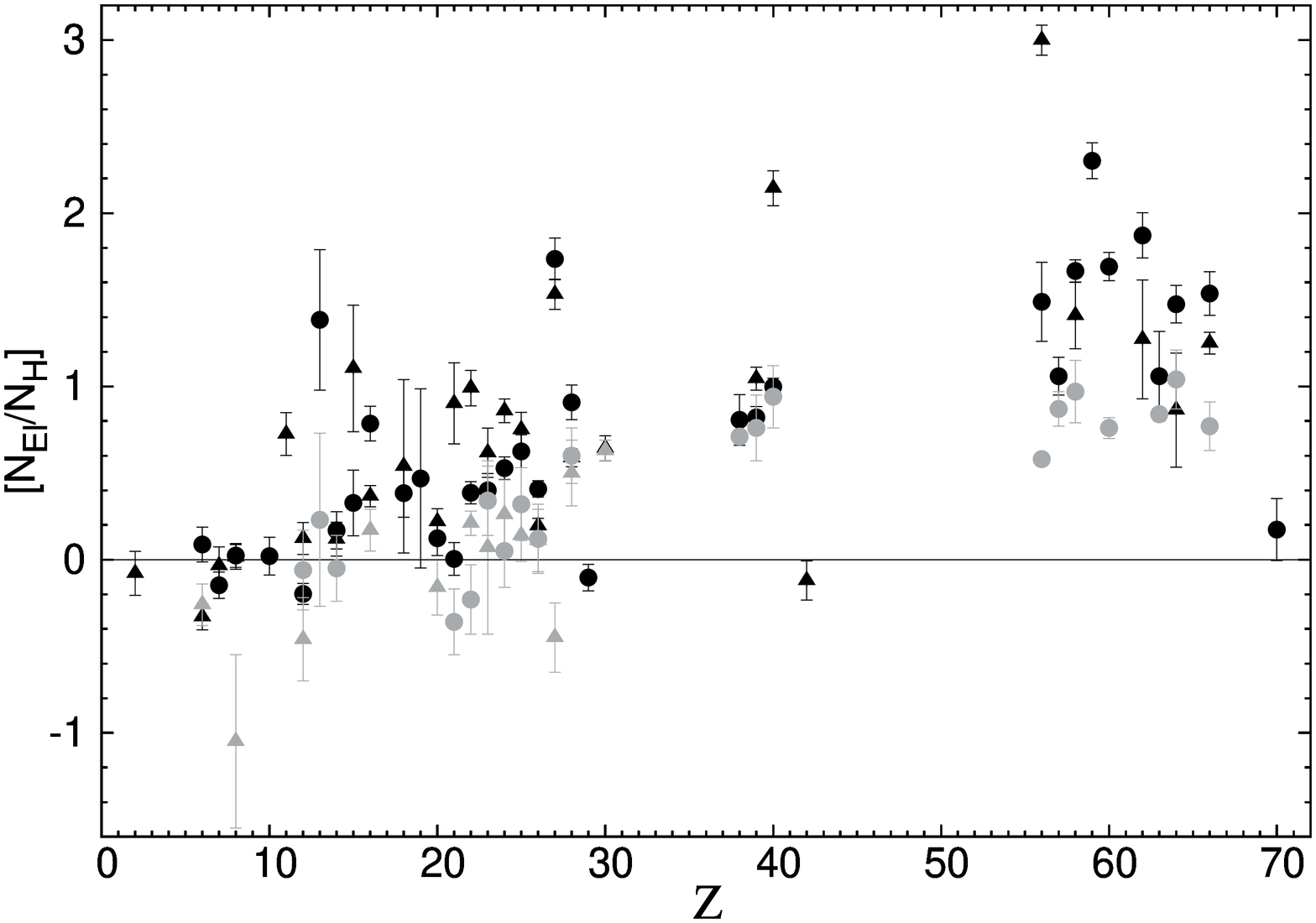} &
\includegraphics[width=2.3in,angle=-90]{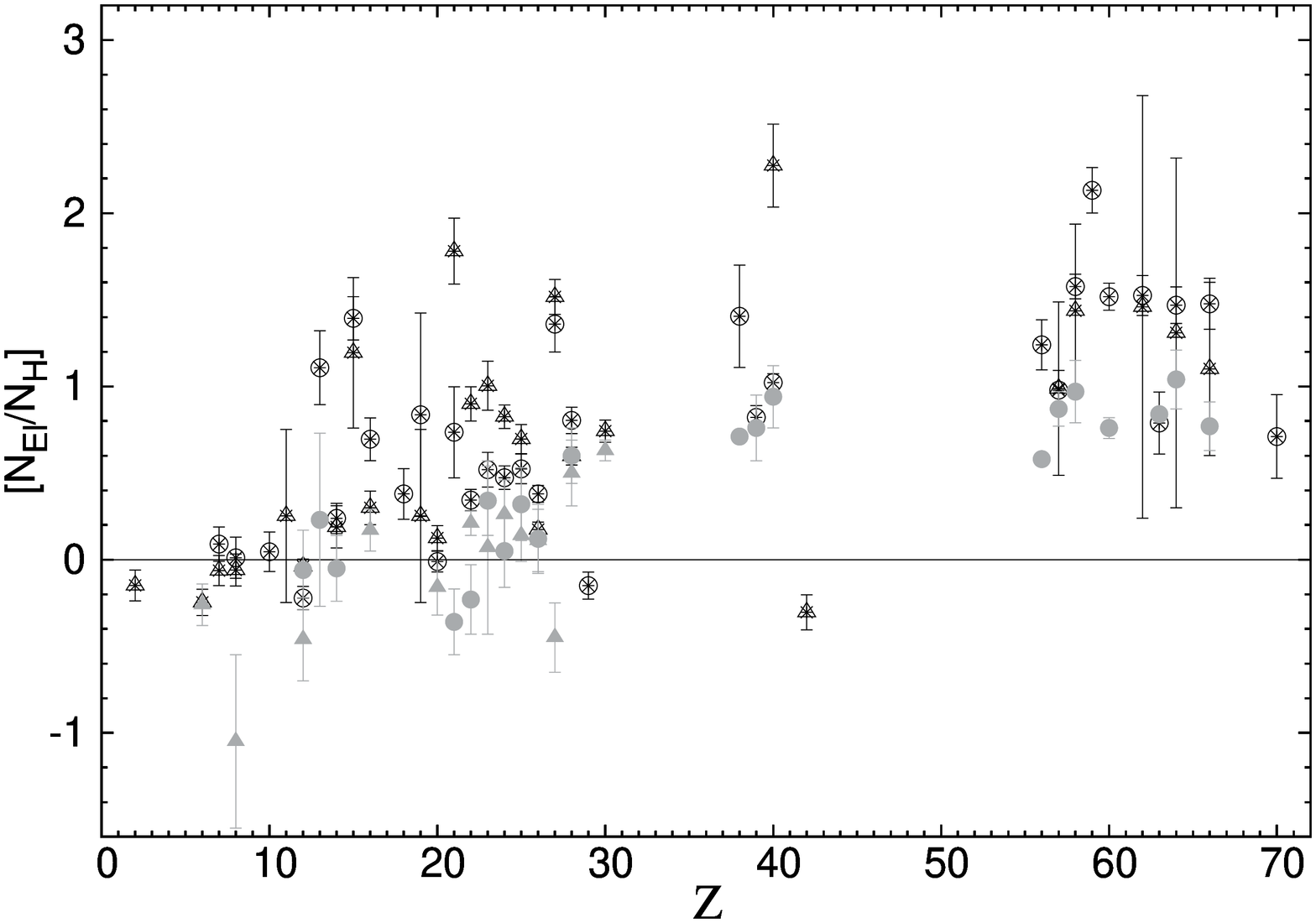} \\
\end{tabular}
\caption{ Average abundance of neutral (triangles) and once ionised ions (circles) in stellar atmosphere of HD~157087 with respect to their solar abundance (dashed line). The comparison of the results obtained by \citet{Yuce+11} (shaded symbols) with the data derived from analysis of the combined spectrum created from the two spectra observed on February 10 (filled symbols) is shown at the left part, while the comparison of those results with the data derived from analysis of the spectrum observed on February 15 (open symbols) is presented on the right part. }
\label{fig4}
\end{center}
\end{figure*}

\begin{figure*}
\begin{center}
\begin{tabular}{cc}
\includegraphics[width=2.1in,angle=-90]{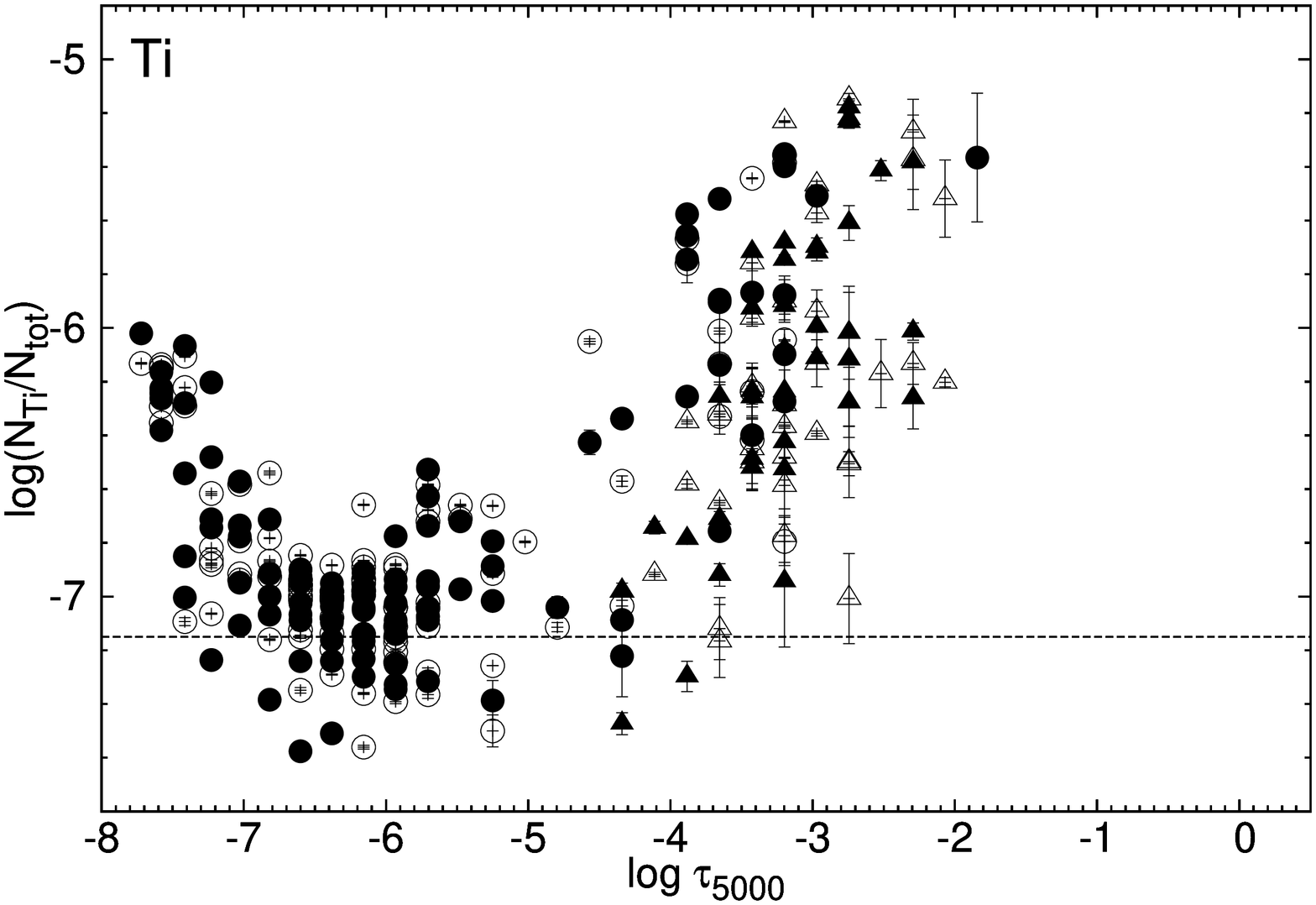} &   
\includegraphics[width=2.1in,angle=-90]{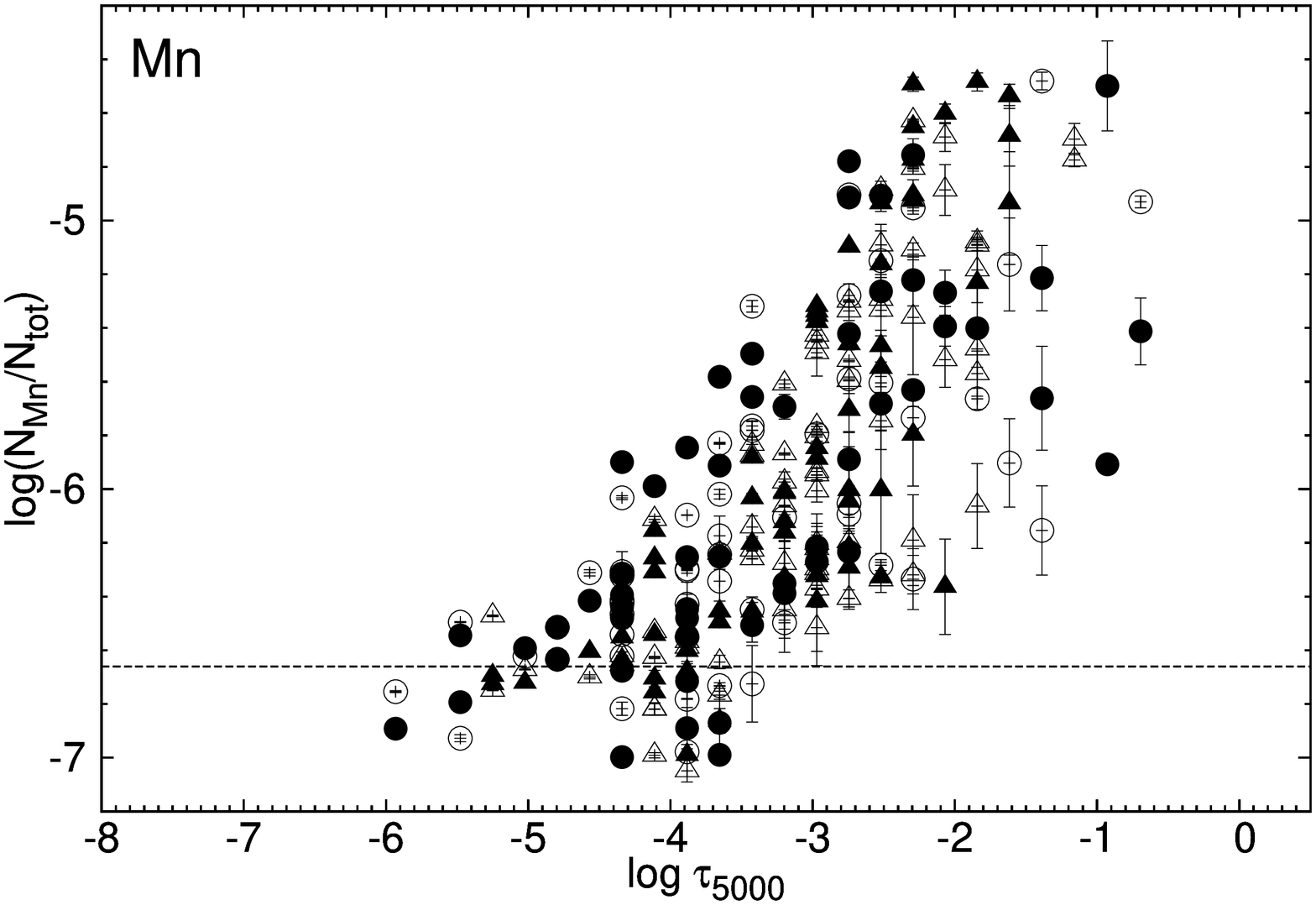}\\   
\includegraphics[width=2.1in,angle=-90]{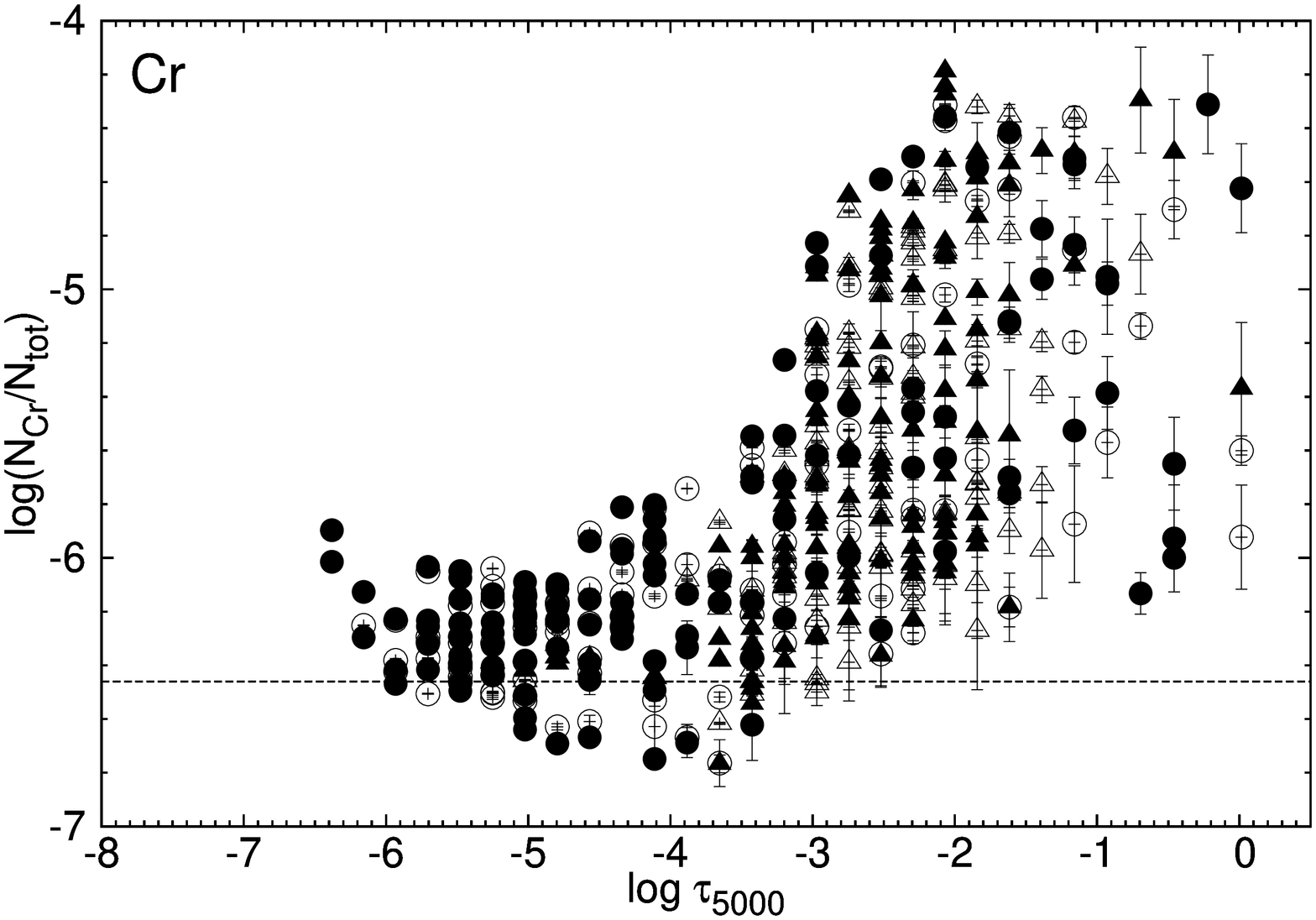} &    
\includegraphics[width=2.1in,angle=-90]{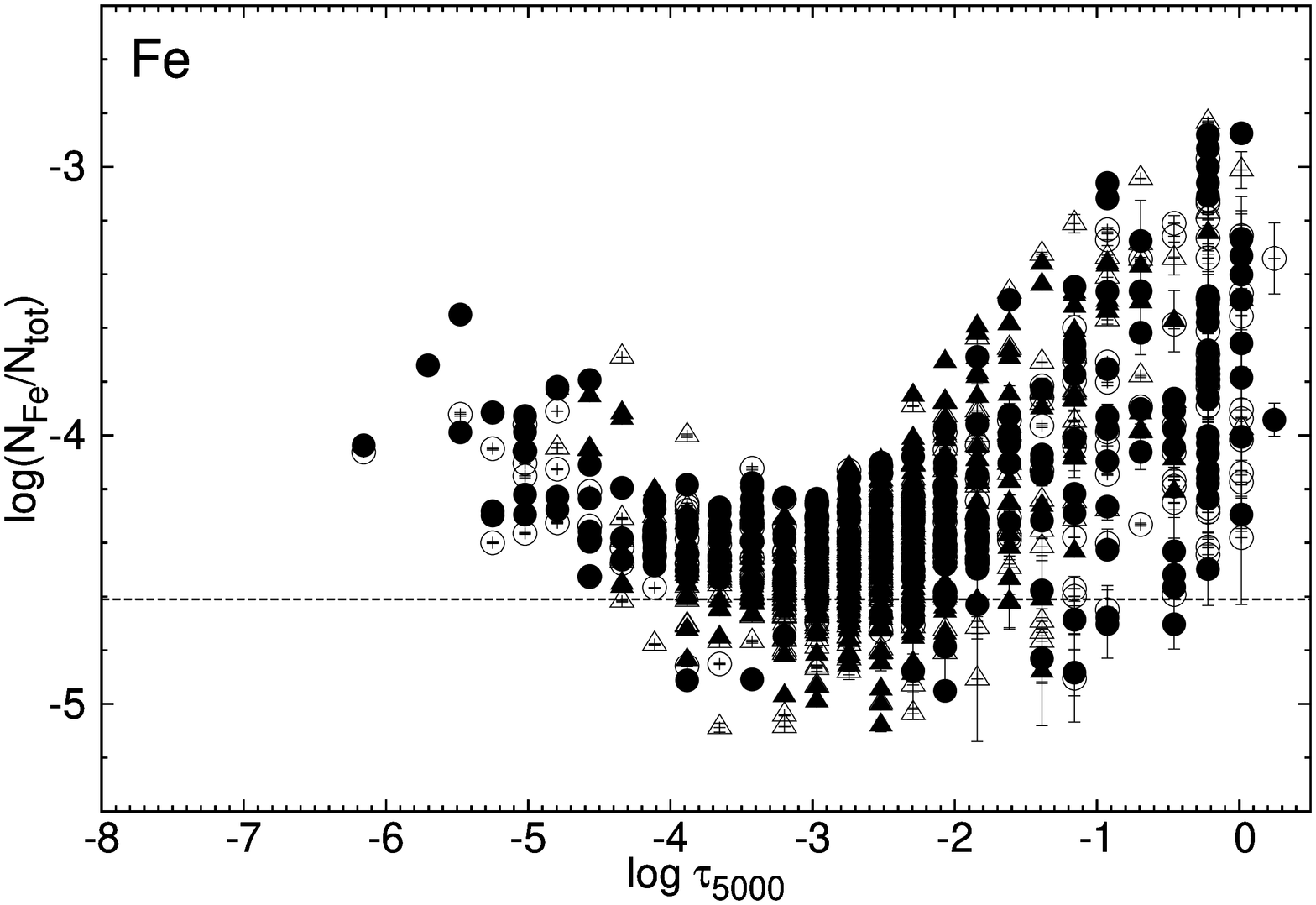} \\   
\includegraphics[width=2.1in,angle=-90]{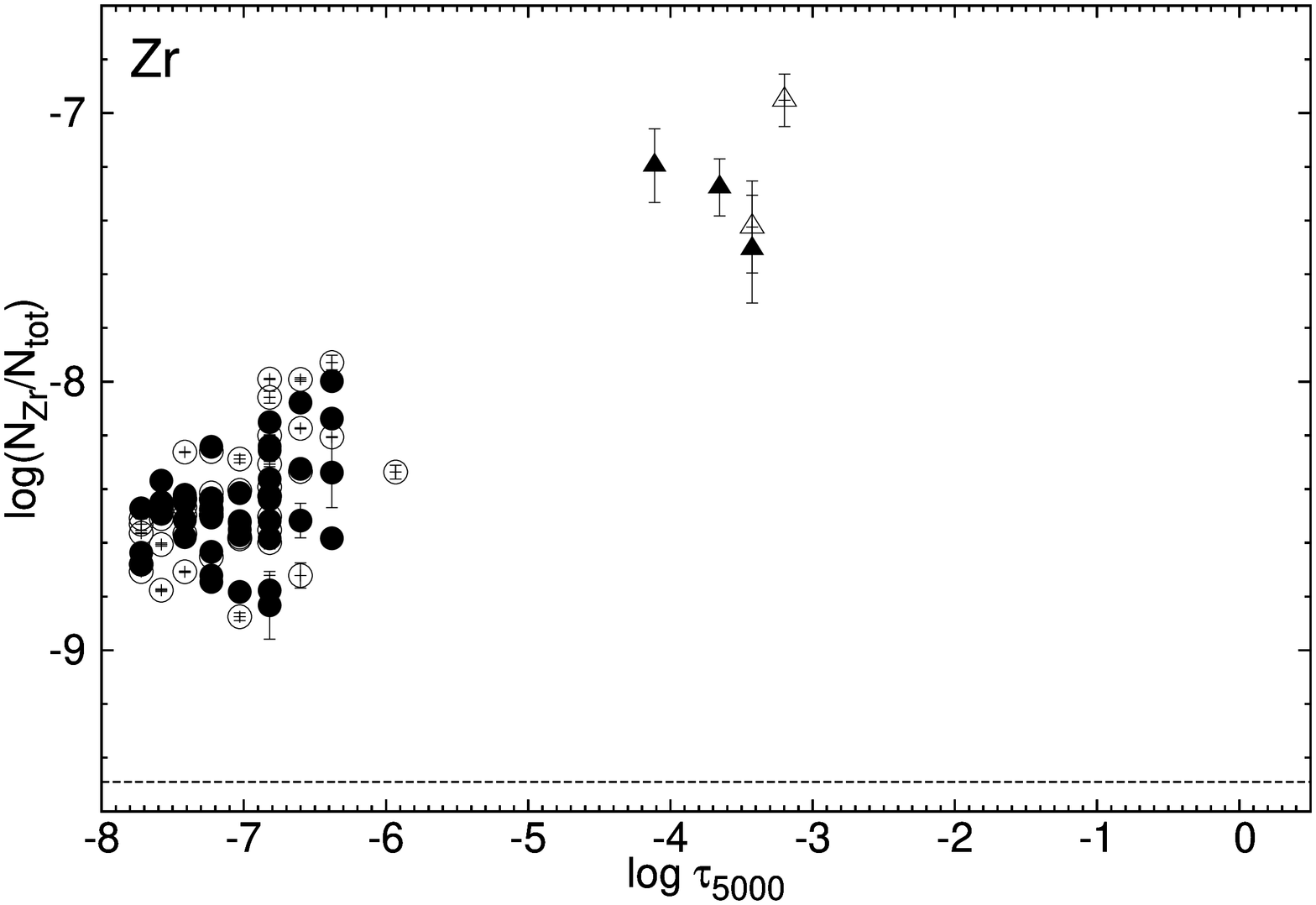} &    
\includegraphics[width=2.1in,angle=-90]{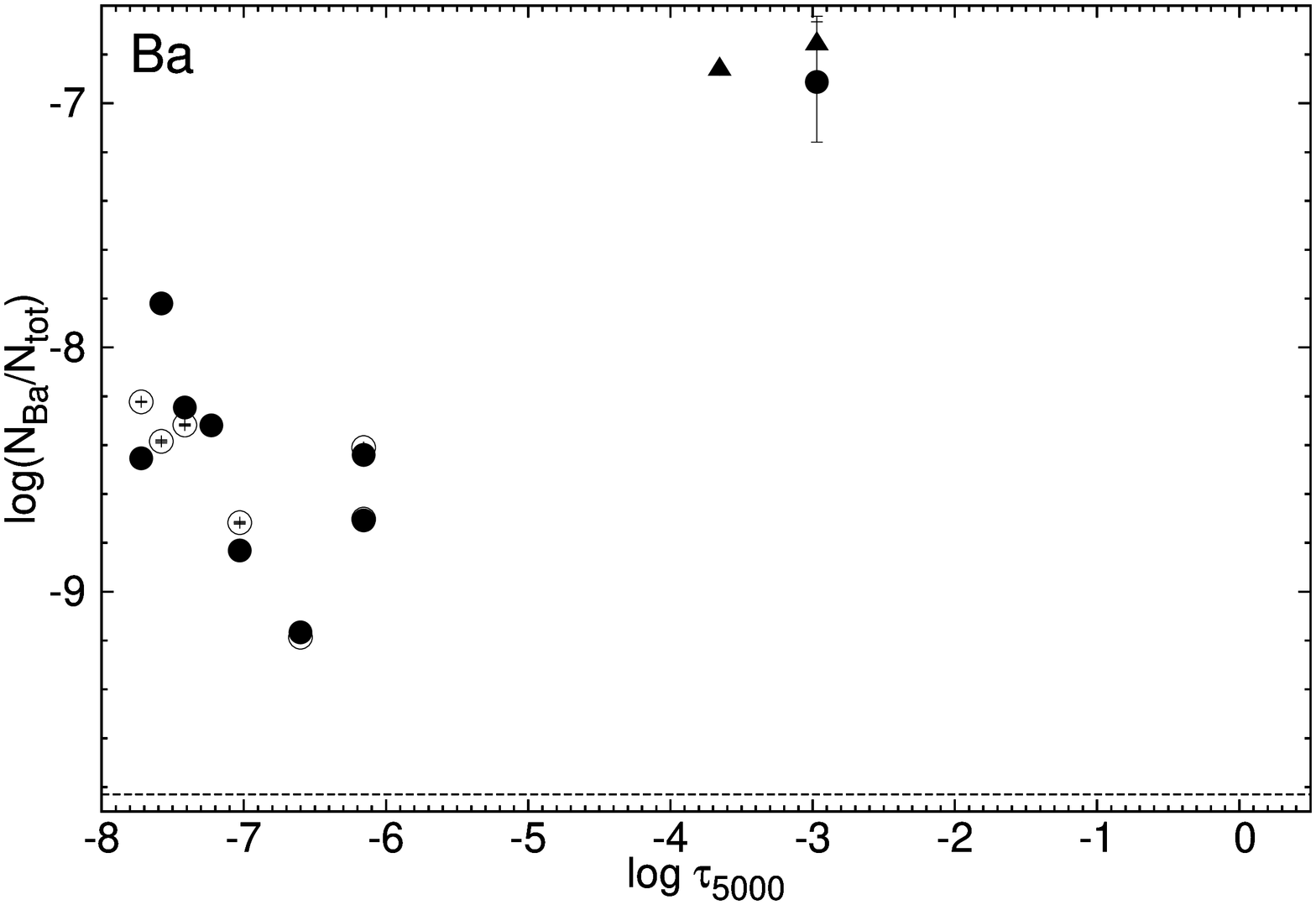} \\   
\includegraphics[width=2.1in,angle=-90]{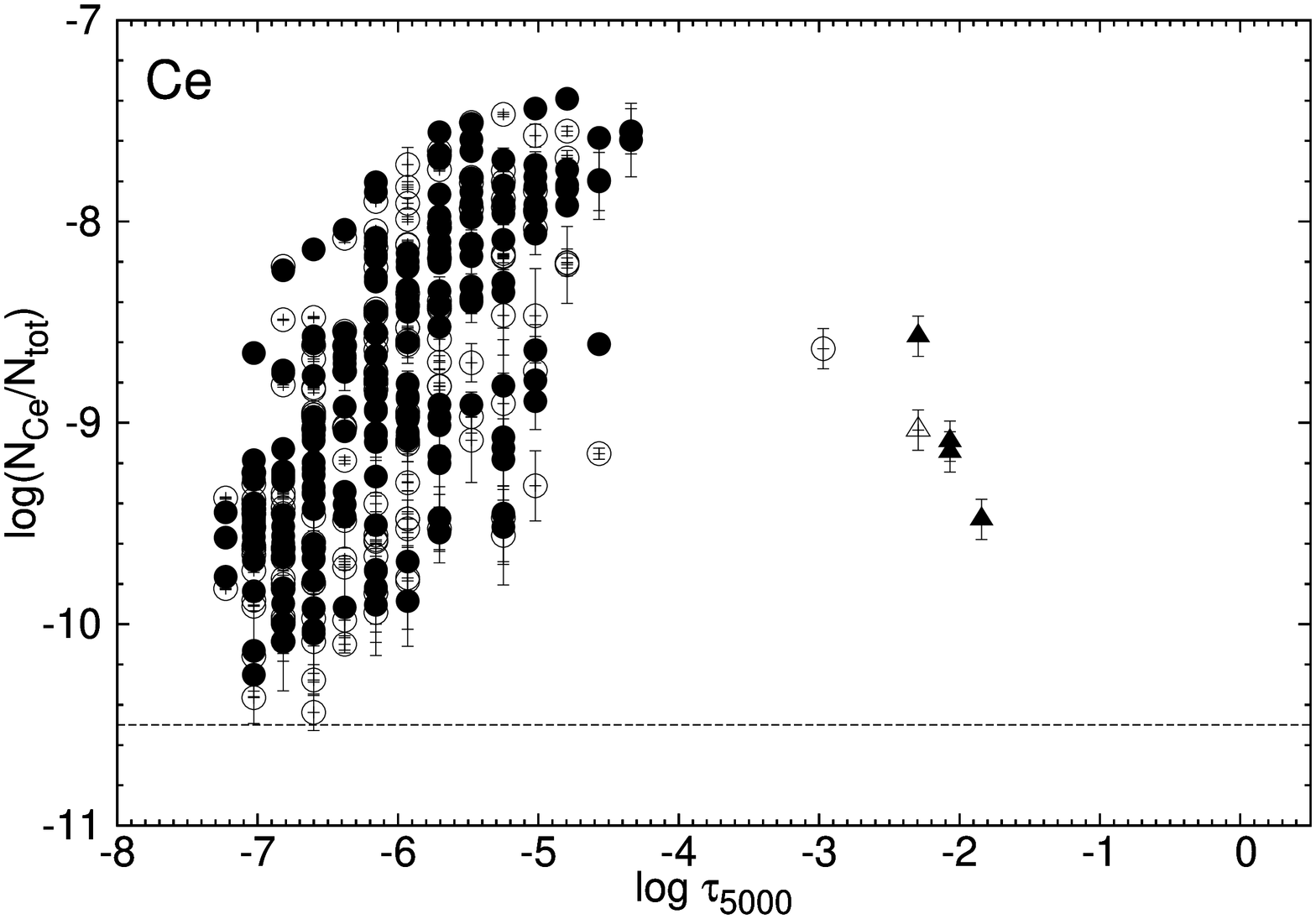} &   
\includegraphics[width=2.1in,angle=-90]{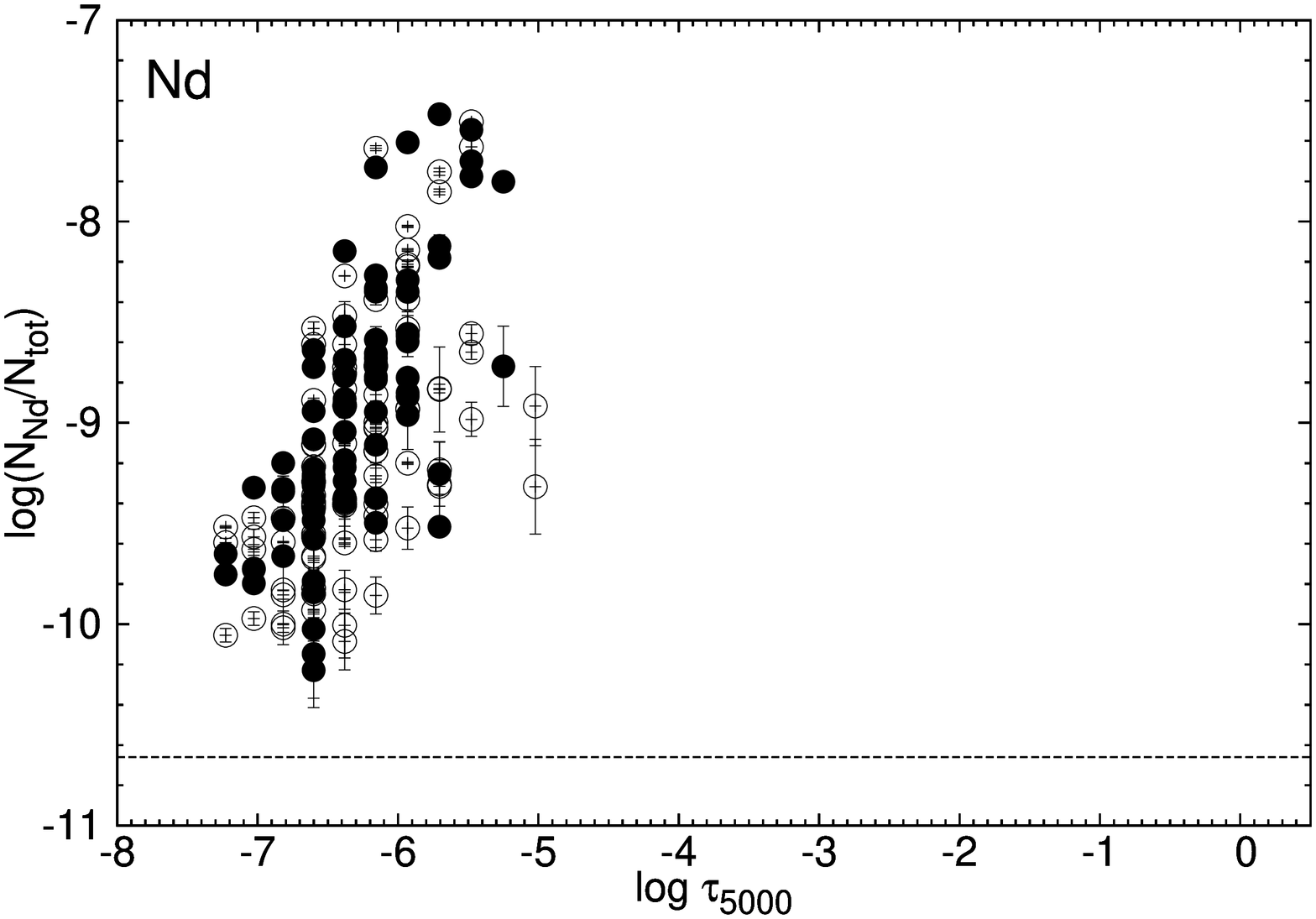}\\
\end{tabular}
\caption{ Distribution of titanium, manganese, chromium, iron, zirconium, barium, cerium and neodymium 
abundance with optical depth in stellar atmosphere of HD~157087.
The filled symbols present results derived from analysis of the combined spectra observed on February 10 (see Table~\ref{tab1}), while the open symbols stand for the data obtained from analysis of the spectrum observed on February 15. The triangles and circles represent the data obtained respectively for the neutral and once ionised atoms.
For each studied element its solar abundance with respect to the total number of atoms in 1 cm$^3$ is specified by a horizontal dashed line. }
\label{fig5}
\end{center}
\end{figure*}

\subsection{Average abundance estimates}
\label{results}

A post-simulation automatic procedure uses the profiles with a ``good" fit obtained from the separate analysis of two spectra to estimated an average abundance for each ion.
Then we compare our estimates of average abundance derived for different ions with the previously published data \citep{Yuce+11}
to verify if the chemical peculiarity of the studied star is variable in time.
Table~\ref{tab5} contains the average estimates of ions abundance derived for HD~157087 (HD~6455) in this study and by \citet{Yuce+11}.
The first column specifies the studied ion, the 2nd and 3rd columns present respectively an average abundance with estimation errors and a number of analysed profiles selected from the combined spectra observed on Feb.~10 (see Table~\ref{tab1}), while the 4th and 5th columns provide the same information for the line profiles selected from the spectrum observed on Feb.~15. The 6th and 7th columns show similar data for the average abundance estimates obtained recently for this star by \citet{Yuce+11} taking into account the new estimates of solar abundance \citep{Grevesse+10,Grevesse+15,Scott+15a,Scott+15b}. The error-bars have been calculated taking into account the estimation errors of stellar average abundance and the precision of solar abundance estimates.

Among the analysed chemical species only  C\,{\sc i}, Mg\,{\sc ii} and Mo\,{\sc i} (Feb.~15) appear to be slightly underabundant ($\sim$ -0.2 -- -0.3 dex), when C\,{\sc ii} and Mg\,{\sc i} seem to have solar abundance in stellar atmosphere of HD~157087. Meanwhile, \citet{Yuce+11} have found solar abundance for C\,{\sc i} and Mg\,{\sc ii}, and an abundance deficit for Mg\,{\sc i}. These authors have also reported underabundance of O\,{\sc i} ($\sim$ -1.0 dex), of Sc\,{\sc ii} and Co\,{\sc i} ($\sim$ -0.4 dex), while we have found solar abundance O\,{\sc i} and O\,{\sc ii}, and enhanced abundance for Sc\,{\sc i}, Sc\,{\sc ii} (Feb.~15) and Co\,{\sc i}.

Our analysis has shown that the abundance of Na\,{\sc i} (Feb.~10), S\,{\sc i}, S\,{\sc ii}, Ar\,{\sc ii}, Sc\,{\sc ii} (Feb.~15), Ti\,{\sc ii}, V\,{\sc i} (Feb.~10), V\,{\sc ii}, Cr\,{\sc ii}, Mn\,{\sc i}, Mn\,{\sc ii}, Fe\,{\sc i}, Fe\,{\sc ii}, Ni\,{\sc i}, Zn\,{\sc i} is significantly enhanced ($\sim$ +0.2 -- +0.7 dex), while Al\,{\sc ii}, P\,{\sc i}, P\,{\sc ii} (Feb.~15), Sc\,{\sc i}, Ti\,{\sc i}, V\,{\sc i} (Feb.~15), Cr\,{\sc i}, Co\,{\sc i}, Co\,{\sc ii}, Ni\,{\sc ii}, Sr\,{\sc ii}, Y\,{\sc i} (Feb.~10), Y\,{\sc ii} and REE (La, Ce, Pr\,{\sc ii}, Nd\,{\sc ii}, Sm, Eu\,{\sc ii}, Gd, Dy) are strongly overabundant ($\sim$ +0.8 -- +1.7 dex) in stellar atmosphere of HD~157087 (see Tab.~\ref{tab5}). In the present study, an extremely high abundance of Zr\,{\sc i} ($\geq$ +2.1 -- +2.3 dex) and Ba\,{\sc i} (Feb.~10: $\sim$ +3.0 dex) has also been found for this star, while Zr\,{\sc ii} and Ba\,{\sc ii} are strongly overabundant ($\sim$ +1.0 -- +1.5 dex) in the two analysed spectra. The derived abundance of Yb\,{\sc ii} is solar in the spectra obtained on Feb.~10 or enhanced in the spectrum obtained on Feb.~15. \citet{Yuce+11} have reported similar abundance estimated for S\,{\sc i}, V\,{\sc ii}, Ni\,{\sc i}, Ni\,{\sc ii}, Zn\,{\sc i}, Y\,{\sc ii}, Zr\,{\sc ii}, La\,{\sc ii}, Ce\,{\sc ii}, Nd\,{\sc ii}, Eu\,{\sc ii}, and Gd\,{\sc ii}, but, in general, we have found higher abundance estimates for the other ions (see Fig.~\ref{fig4}).

Figure~\ref{fig4} shows the derived average abundances of neutral and once ionised ions in stellar atmosphere of HD~157087 with respect to their solar abundance. One can clearly see the significant discrepancies between the abundance estimates derived in this study and those found by \citet{Yuce+11}. Taking into account the time difference when the reported spectra and the spectra of \citet{Yuce+11} have been obtained, the detected time variation of abundance estimates for different ions can indicate that this star is a member of binary system (see Sec.~\ref{binary}).

\begin{table}
\begin{center}
\caption{Slopes of abundance change with respect to optical depths.}
\label{tab6}
\begin{tabular}{lcrr}\hline
Element & $\log\tau_{5000}$ & \multicolumn{2}{c|}{Slope} \\
        & region            & Feb. 10$^a$ & Feb. 15$^a$ \\
\hline
C  &  -4.0 -- 0.5  &  0.24$\pm$0.05 &  0.34$\pm$0.11 \\
N  &  -3.5 -- 0.5  &  0.04$\pm$0.05 &  0.09$\pm$0.04 \\ 
O  &  -3.5 -- 0.5  &  0.00$\pm$0.04 &  0.00$\pm$0.09 \\ 
Mg &  -6.0 -- -1.0 & -0.15$\pm$0.06 & -0.03$\pm$0.06 \\
Si &  -5.0 -- 0.5  &  0.09$\pm$0.05 &  0.15$\pm$0.05 \\ 
P  &  -3.0 -- 0.5  & -0.50$\pm$0.11 &  0.30$\pm$0.31 \\
S  &  -3.5 -- 0.5  &  0.22$\pm$0.04 &  0.20$\pm$0.07 \\
Ca &  -4.0 -- 0.5  &  0.47$\pm$0.04 &  0.36$\pm$0.07 \\
Sc &  -8.0 -- -2.0 &  0.25$\pm$0.04 &  0.41$\pm$0.06 \\
Ti &  -8.0 -- -6.0 & -0.54$\pm$0.06 & -0.51$\pm$0.06 \\
   &  -6.0 -- -1.5 &  0.42$\pm$0.04 &  0.35$\pm$0.04 \\
V  &  -7.5 -- -2.5 &  0.18$\pm$0.04 &  0.23$\pm$0.03 \\
Cr &  -4.0 -- 0.5  &  0.39$\pm$0.04 &  0.37$\pm$0.05 \\
Mn &  -6.0 -- -0.5 &  0.50$\pm$0.04 &  0.46$\pm$0.04 \\
Fe &  -8.0 -- -3.2 & -0.28$\pm$0.03 & -0.23$\pm$0.03 \\
   &  -3.2 -- 0.5  &  0.29$\pm$0.01 &  0.31$\pm$0.01 \\
Co &  -5.5 -- 0.0  &  0.49$\pm$0.06 &  0.53$\pm$0.07 \\
Ni &  -6.0 -- 0.5  &  0.22$\pm$0.03 &  0.19$\pm$0.03 \\
Y  &  -8.0 -- -0.5 & -0.02$\pm$0.01 & -0.04$\pm$0.02 \\ 
Zr &  -8.0 -- -3.0 &  0.31$\pm$0.03 &  0.32$\pm$0.03 \\
Ba &  -8.0 -- -6.5 & -0.95$\pm$0.33 & -0.85$\pm$0.11 \\
   &  -6.5 -- -2.0 &  0.57$\pm$0.07 & -------------- \\
Ce &  -8.0 -- -4.0 &  0.72$\pm$0.05 &  0.76$\pm$0.07 \\ 
Dy &  -8.0 -- -5.5 &  0.64$\pm$0.26 &  1.19$\pm$0.20 \\ 
Gd &  -8.0 -- -5.5 &  1.12$\pm$0.16 &  0.80$\pm$0.24 \\ 
La &  -8.0 -- -6.0 &  0.89$\pm$0.20 &  0.81$\pm$0.21 \\ 
Nd &  -8.0 -- -4.5 &  1.04$\pm$0.12 &  0.74$\pm$0.12 \\ 
Pr &  -8.0 -- -4.5 &  0.91$\pm$0.14 &  1.01$\pm$0.22 \\ 
Sm &  -8.0 -- -5.0 &  0.98$\pm$0.23 &  1.14$\pm$0.19 \\ 
Yb &  -7.0 -- -2.5 & -0.59$\pm$0.15 & -0.19$\pm$0.14 \\ 
\hline
\end{tabular}
\end{center}
{\it Notes:} $^a$results for two spectra observed on different nights (see Table~\ref{tab1}).
\end{table}

\subsection{Analysis of vertical stratification of element abundance}
\label{strat}

Analysis of the variation of elements abundance with the atmospheric depth (taking into account their neutral and once ionised ions) reveals evidences of vertical abundance stratification of some chemical species in stellar atmosphere of HD~157087 (see Figs.~\ref{fig5}). 
Table~\ref{tab6} presents the slopes derived from the linear regression of abundance change with optical depths found in the spectra obtained on 10 Feb. (3-rd column) and on 15 Feb. (4-th column). The slopes of linear regression have been calculated for different regions of optical depths shown at the second column. If for some chemical element its abundance estimates are obtained for a very narrow range of optical depth one can not properly determine its abundance change with optical depth. Therefore, in table~\ref{tab6} the data for He, Ne, Na, Al, Ar, K, Cu, Zn, Sr, Mo and Eu are not shown because all their abundance estimates occupy a very narrow range of optical depth.

Some elements, like C, S, Ca, Sc, V, Cr, Mn, Co, Ni and Zr show significant trends of abundance increase toward the deeper atmospheric depths. For some of them the abundance increase reaches up to 2 dex (see Fig.~\ref{fig5}). Abundance of titanium, iron and probably barium reaches its minimum respectively at the optical depth $\log\tau_{5000}$= -6.5 -- -6.0, $\log\tau_{5000}$= -3.5 -- 2.8, $\log\tau_{5000}$= -6.8 -- -6.3, and has a tendency to increase towards the upper and deeper atmospheric layers. For barium this picture is less convincing because of much smaller number of abundance estimates that prove this behaviour (see Fig.~\ref{fig5}).

Other elements, like N, O, Si, P (Feb.~15) and Y, do not show any significant stratification of their vertical abundance. Meanwhile, results of abundance analysis of spectra obtained on Feb.~10 indicate that the phosphore increase its abundance towards the upper levels of stellar atmosphere. The magnesium (Feb.~10) seems to slightly increase its abundance toward the upper atmospheric layers in HD~157087, but the derived slope of abundance change with optical depths is not significant (see Tab.~\ref{tab6}).

REE (Ce, Dy, Gd, La, Nd, Pr, Sm and Yb) are significantly enhanced at the layers close to the stellar surface and show same increase of their abundance towards the deeper atmospheric layers. Nevertheless, the range of optical depths, where one can trace an abundance increase, is rather narrow. Usually the range is $\log\tau_{5000}$= -7.5 -- -5.0 (see Fig.~\ref{fig5} and Tab.~\ref{tab6}). In the deeper layers of stellar atmosphere, some REE (for example, cerium) do show an abundance similar to the one (or smaller than the one) found at the layers close to the stellar surface.

We can find similar situation for the vertical stratification of zirconium abundance based on the analysis of only Zr\,{\sc ii} line profiles taking into account a rather narrow range of optical depths of their formation. Nevertheless, the analysis of Zr\,{\sc i} line profiles results in even higher abundance estimates for the deeper atmospheric layers (see Fig.~\ref{fig5}). Therefore, in this case we have a stronger evidence of the vertical abundance stratification for the zirconium.

\begin{table}
\begin{center}
\caption{Radial velocity measurements.}
\label{tab7}
\begin{tabular}{lccc}\hline
Heliocentric& \multicolumn{2}{c|}{$V_{\rm r}$, km s$^{-1}$ } & Source \\
Julian Date & Primary  & Secondary & \\
\hline
21741.885 &  -2.8$\pm$ 1.5 &  & \citet{Plaskett+22} \\
21759.868 &   0.3$\pm$ 1.5 &  & --- \\
21780.743 &  -1.1$\pm$ 1.5 &  & --- \\
22181.693 &  -4.9$\pm$ 1.5 &  & --- \\
22496.833 & -10.3$\pm$ 1.5 &  & --- \\
49552.793 &  -3.6$\pm$ 0.2 &  & \citet{Yuce+11} \\
50555.037 &   1.6$\pm$ 0.1 &  & --- \\
50935.013 &   5.0$\pm$ 0.2 &  & --- \\
51399.843 &   1.3$\pm$ 0.8 &  & --- \\
51449.669 &   1.6$\pm$ 0.6 &  & --- \\
52103.786 &   3.1$\pm$ 0.6 &  & --- \\
52103.803 &   2.8$\pm$ 0.7 &  & --- \\
52487.708 &   3.0$\pm$ 0.3 &  & --- \\
53174.832 &   1.3$\pm$ 0.3 &  & --- \\
53176.888 &   0.6$\pm$ 0.3 &  & --- \\
53176.971 &   1.0$\pm$ 0.3 &  & --- \\
53177.782 &   1.0$\pm$ 0.1 &  & --- \\
53215.729 &   1.4$\pm$ 0.2 &  & --- \\
56699.129 &  -9.64$\pm$ 0.05&  & this study \\
56704.085 &  -9.55$\pm$ 0.05&  & this study \\
\hline
\end{tabular}
\end{center}
\end{table}

\section{Binary system}
\label{binary}

Detection of vertical abundance stratification for some metals whose average abundance changes with time (see Tables~\ref{tab5},~\ref{tab6}) does not agree well with a hypothesis that
HD~157087 is a singular star of Am type. The detected change of average abundance can be explained in terms of binary nature of this star, as it was reported early by \citet{Bidelman88}.
In a binary system, the secondary component may contribute significantly to the strength of observed line profiles and cause a variability of chemical abundance with orbital phase \citep{Lyubimkov89}. The analysis of line profiles in the available spectra of HD~157087 does not show clear signatures (line profiles) that can be attributed to the secondary component.
Meanwhile, \citet{Makarov+Kaplan05} have reported HD~157087 as an astrometric binary system that shows some changes in its short-term proper motion. Usually such changes can by detected only in the long-periodic binaries with the periods higher than 6 years \citep{Makarov+Kaplan05}.

The estimates of radial velocity derived in this study are shown in Table~\ref{tab7} together with the previously published data of \citet{Plaskett+22} and \citet{Yuce+11}. These data have been analysed with the help of version 1.2 of the code PERIOD04 developed by \citet{Lenz+Breger05}, which was specially designed to carry out a statistical analysis of large astronomical data sets with significant gaps. The Fourier transformation of the available data on radial velocity results in rather spurious peaks in the frequency diagram. The most significant peaks have been found one by one and pre-whitened from the observed data using the least-squares fitting procedure \citep{Lenz+Breger05}. The analysis results in the detection of five periods with the amplitudes having a significant signal-to-noise ratio at the detected frequency (see Table~\ref{tab8}).
Finally, the available radial velocity data were fitted taking into account all five frequencies using the least-squares procedure and the obtained periods, amplitudes and phases are shown respectively in the first, second and third columns of the Table~\ref{tab8}. The presented uncertainties are the highest ones obtained from the error-matrix produced by the Levenberg-Marquardt non-linear least-squares fitting procedure and from the
Monte Carlo simulation of a signal with the detected frequencies \citep{Lenz+Breger05}. They do not take into account the estimation errors reported for the {$V_{\rm r}$ measurements. 
In this approach the uncorrelated uncertainties for the frequency and the phase have been derived \citep{Montgomery+Odonoghue99}. The 4th column of the Table~\ref{tab8} contains signal-to-noise ratio calculated for each derived frequency from simulation of the white noise using the dispersion of residual data left after the pre-whitening procedure \citep{Lenz+Breger05}.
Taking into account the quite high uncertainties for the detected periods and their amplitudes and phases (see Table~\ref{tab8}) they can only be considered as indicators of the long term and the short term periodic variations of {$V_{\rm r}$. To determine these periods with a higher precision one needs to acquire additional spectra of HD~157087 and to derive new data for radial velocity.

\begin{table}
\begin{center}
\caption{Periodic variations of radial velocity.}
\label{tab8}
\begin{tabular}{cccr}\hline
 Period, & Amplitude, & Phase & S/N \\
   d  & km s$^{-1}$ &  & \\
\hline 
8469.6$\pm$2477.1 & 4.55$\pm$1.93 & 0.050$\pm$0.238 & 476 \\
1710.4$\pm$119.6 & 2.32$\pm$1.16 & 0.563$\pm$0.196 & 243 \\
4.081$\pm$0.012 & 1.66$\pm$1.17 & 0.003$\pm$0.282 & 161 \\
5.874$\pm$0.093 & 0.99$\pm$1.49 & 0.504$\pm$0.335 & 98 \\
3.144$\pm$0.004 & 0.43$\pm$1.26 & 0.404$\pm$0.332 & 42 \\
\hline
\end{tabular}
\end{center}
\end{table}

Considering that the previous measurements of radial velocity are poor constraining and are not well folded in the time domain \citep{Plaskett+22,Yuce+11} the derived periods may differ significantly from the real ones. Nevertheless, in Table~\ref{tab8} one can clearly identify a presence of long periodic (several years) variability of radial velocity that can be explained by orbital motion in a binary system. This fact supports the hypothesis of \citet{Makarov+Kaplan05} that HD~157087 is an astrometric binary system with a period higher than 6 years. On the other hand, the short periodic variations (3 -- 6 days) of radial velocity are also evident and can be caused by an axial rotation of the star with abundance inhomogeneities on its surface or by
a short periodic binary that rotates around a third star (assuming that HD~157087 is a triple system).
In any case, to measure the short period of radial velocity variability in HD~157087 with higher precision one needs to acquire additional spectra of this star that are well folded over the period of 3 -- 6 days.

\section{Discussion}
\label{discus}

\citet{Renson+Manfroid09} have considered HD~157087 as a candidate to CP stars having spectral class A2, while \citet{Yuce+11} have reported HD~157087 as a marginal Am basing on their results of abundance analysis. \citet{Bidelman88} has also mentioned HD~157087 as a marginal Am star which probably is a member of a binary system, while \citet{Makarov+Kaplan05} have shown that HD~157087 is member of an astrometric binary system. Present paper aims to study the abundance stratification of chemical elements in the stellar atmosphere of HD~157087 and to analyse a change of its average abundance with time to verify the nature of this object.

The effective temperature derived in this study from the fitting of Balmer line profiles \citep{Khalack+LeBlanc15a} 
is in good accordance with the estimate of $T_{\rm eff}$ obtained from the two different photometric temperature calibrations \citep{Napiwotzky+93,Netopil+08}. 
Taking into account the estimation errors, the derived here values of $T_{\rm eff}$ and $\log{g}$ (see Table~\ref{tab3}) agree relatively well with $T_{\rm eff}$ = 8700$\pm$150~K and $\log{g}$ = 3.25$\pm$0.20 obtained for HD~157087 by \citet{Yuce+11}. Basing on the visual magnitude and parallax provided for this object by Hipparcos catalogue \citep{ESA97,vanLeeuwen07} we have derived its luminosity and found its positions at the HR diagram (see Fig.~\ref{fig2}).
Apparently, for the derived effective temperature and luminosity the interpolation of isochrones calculated by \citet{Bressan+12} for the metallicity [M/H]=0.0 results in two sets of age, mass, surface gravity and stellar radius (see Table~\ref{tab3}). The estimates of stellar age (mass, surface gravity and radius) obtained for HD~157087 in the two sets are similar one to another taking into account their uncertainties. The value of $\log{g}$ obtained from the interpolation of isochrones is very close to the one derived from the fitting of Balmer line profiles. Our estimate of stellar radius also agrees well with the $R_{\rm *}$ = 4.5 $R_{\rm \odot}$ provided for this object by the Catalogue of Apparent Diameters and Absolute Radii \citep{Pasinetti-Fracassini+01}. It is found that HD~157087 does not posses a detectable magnetic field (see Table~\ref{tab1}).

The stellar atmosphere model has been calculated with the help of PHOENIX15 code \citep{Hauschildt+97} for $T_{\rm eff}$ = 8882~K, $\log{g}$ = 3.57, [M/H]=0.0 assuming a microturbulence velocity $\xi$ = 2.1 km s$^{-1}$ \citep{Yuce+11}. It was used to carry out the abundance analysis employing the code ZEEMAN2 \citep{Landstreet88,Khalack+Wade06,Khalack+07} under the assumption that HD~157087 is a singular star with no magnetic field. From the analysis of 929 profiles in the combined spectrum obtained on Feb.~10 and of 746 line profiles in the spectrum obtained on Feb.~15, we have estimated its radial velocity (see Table~\ref{tab7}), $V\sin(i)$= 10.35$\pm$0.05 km s$^{-1}$ (Feb.~10) and 10.18$\pm$0.05 km s$^{-1}$ (Feb.~15), and average abundance of chemical elements (see Table~\ref{tab5}).
The average abundances of chemical elements derived here usually are higher (especially for the REE) than the data reported by \citet{Yuce+11} for the studied star (see Fig.~\ref{fig4}). This fact argues in favour of the binary nature of HD~157087. Our analysis of the available measurements of radial velocity results in detection of long periodic (5 -- 23 years) and short periodic (3 -- 6 days) variations (see Table~\ref{tab8}). The long periodic variation agrees well with the idea that HD~157087 is the astrometric binary system with a period higher than 6 years \citep{Makarov+Kaplan05}. Meanwhile, the presence of short periodic variation of $V_{\rm r}$ and the changes of average abundance with time suggest that HD~157087 may in reality be a triple system in which a short periodic binary rotates around a third star.

The convective stellar atmosphere of an Am star should posses a more or less uniform abundance everywhere at the stellar surface and at different optical depths due to the turbulent motions that rapidly smooth out any abundance inhomogeneities \citep{Michaud+83}. According to \citet{Preston74} and \citet{Adelman87} stellar atmospheres of classical Am stars usually are rich in heavy elements and shows some deficit of calcium and scandium.
The results of abundance analysis found in this study argue against the classification of HD~157087 as even a marginal Am star \citep{Bidelman88,Yuce+11}, because stellar atmosphere of this star shows solar calcium abundance (Ca\,{\sc i} seems to be enhanced in the spectrum obtained on 10 Feb.) and scandium appears to be significantly overabundant. Analysis of vertical abundance stratification of elements in the stellar atmosphere of HD~157087 reveals that C, S, Ca, Sc, V, Cr, Mn, Co, Ni and Zr show significant trends of abundance increase toward the deeper atmospheric layers (see Table~\ref{tab6}). This fact does not support the classification of HD~157087 as an Am star either, because the Am stars should not posses a vertical abundance stratification of chemical elements in their atmosphere \citep{Michaud+83}. Nevertheless, taking into account the hypothesis of a triple system with the short periodic binary as its member, the reported facts can be explained in terms of assumption that this short periodic binary consist of slowly rotating Am and A stars (or even Ap star with weak magnetic field) which have a similar $T_{\rm eff}$ and $\log{g}$, but different abundance peculiarities. According to \citet{Lyubimkov89} their orbital motion in the short periodic binary can explain the observed variation of average abundance with time. This hypothesis also explains the detection of vertical abundance stratification for several chemical species in HD~157087 at certain orbital phases.

The analysed here spectra have been acquired with ESPaDOnS during a time span of five days.  The estimates of average abundance for the studied chemical species (see Table~\ref{tab5}) and the profiles of vertical abundance stratification (see Table~\ref{tab6}) do not change significantly during this period of time, though the value of $N_{\rm tot}$ obtained for HD~157087 is decreased by 0.008 dex. Therefore, the short period variation of radial velocity might also be attributed to the period of axial rotation of HD~157087. Taking into account that HD~157087 is a member of the astrometric binary system \citep{Makarov+Kaplan05} whose components should be visually separated, it would be difficult to explain under an assumption of only one rotating star the detected long term change of the average abundance over five days (see Table~\ref{tab5}) and during a longer period of time (see Section~\ref{binary}).

To confirm or reject these hypotheses one needs to perform a thorough abundance analysis of additional high resolution and high signal-to-noise spectra of HD~157087 that densely cover the time span in 3 --6 days or even a longer period. This approach will allow one to obtain more measurements of radial velocity that can contribute to the estimation of short variability periods with higher precision. The variability of the average abundance estimates and probably of the vertical abundance stratification of chemical species with time can be studied in more details as well.

\section*{Acknowledgments}

V.K. is grateful to Prof. F. Castelly for the useful comments and helpful advice that led to a significant improvement of this study.
Author acknowledges support from the Natural Sciences and Engineering Research Council of Canada 
and is thankful to the Facult\'{e} des \'{E}tudes Sup\'{e}rieures et de la Recherch and to the Facult\'{e} des Sciences de l'Universit\'{e} de Moncton for the financial support of this research. Part of calculations have been done on the supercomputer {\it briarree} of the university of Montreal, under the guidance of Calcul Qu\'{e}bec and Calcul Canada. The use of this supercomputer is funded by the Canadian Foundation for Innovation (CFI), NanoQu\'{e}bec, RMGA and Research Fund of Qu\'{e}bec - Nature and Technology (FRQNT).
This paper has been typeset from a \TeX/\LaTeX\, file prepared by the author.

\label{lastpage}

\end{document}